\documentclass{aa} % for the abstract without structuration 
\usepackage{graphicx}
\usepackage{txfonts}
 \usepackage[T1]{fontenc}
\usepackage{color}
\usepackage[]{natbib}
\usepackage[switch]{lineno}

        \newcommand{\arc}{$^{\prime\prime}$}
        \newcommand{\msun}{\ensuremath{\mathrm{M}_{\odot}}}
        \newcommand{\lsun}{L$_{\odot}$}
        \newcommand{\sfr}{M$_{\odot}$ yr$^{-1}$}
        
        \newcommand{\kms}{${\rm km/s}$}
        \newcommand{\oiii}{[O{\sc iii}]}
        \newcommand{\cii}{[C{\sc ii}]}
        \newcommand{\lcii}{${\rm L_{\rm [C\textsc{ii}]} }$}
        \newcommand{\loiii}{${\rm L_{\rm [O\textsc{iii}]} }$}
        \newcommand{\mgas}{${\rm M_{\rm gas} }$}

	\newcommand{\ujy}{$\mu$Jy}
	
	\newcommand{\lya}{${\rm Ly\alpha}$}
	\newcommand{\bdf}{${\rm BDF-3299}$}

	\defcitealias{Vallini:2016}{V16}

\begin{document}
%%%%%%%%%%%%%%%%%%%%%%%%%%%%%%%%%%%%%%%%%%%%%%%%%%

%%%%%%%%%%%%%%%%%%% TITLE PAGE %%%%%%%%%%%%%%%%%%%

\title{Extended ionised and clumpy gas in a normal galaxy at z=7.1 revealed by ALMA}

\author{S.~Carniani,\inst{1,2}
R.~Maiolino\inst{1,2},
A.~Pallottini\inst{1,2,3},
L.~Vallini\inst{4},
L.~Pentericci\inst{5},
A.~Ferrara\inst{3},
M.~Castellano\inst{5},
E.~Vanzella\inst{6},
A.~Grazian\inst{5}, 
S.~Gallerani\inst{3},
P.~Santini\inst{5}, 
J.~Wagg\inst{7}, 
A.~Fontana\inst{5}
}
%please confirm you're interested

% List of institutions
\institute{Cavendish Laboratory, University of Cambridge, 19 J. J. Thomson Ave., Cambridge CB3 0HE, UK
\and Kavli Institute for Cosmology, University of Cambridge, Madingley Road, Cambridge CB3 0HA, UK
\and Scuola Normale Superiore, Piazza dei Cavalieri 7, 56126 Pisa, Italy
\and Nordita, KTH Royal Institute of Technology and Stockholm University
Roslagstullsbacken 23, SE-106 91 Stockholm, Sweden
\and INAF - Osservatorio Astronomico di Roma, via Frascati 33, 00040 Monteporzio, Italy
\and INAF - Bologna Astronomical Observatory, via Ranzani 1, I-40127 Bologna, Italy
\and Square Kilometre Array Organization, Jodrell Bank Observatory, Lower Withington, Macclesfield, Cheshire
SK11 9DL, UK}

%\linenumbers

\abstract{We present new ALMA observations of the \oiii88$\mu$m line and high angular resolution observations of the
\cii158$\mu$m line in a normal star forming galaxy at z$=$7.1. 
Previous \cii\ observations of this galaxy
had detected \cii\ emission consistent with the Ly$\alpha$ redshift but spatially slightly
offset relative to the
optical (UV-rest frame) emission.
The new \cii\ observations reveal that the \cii\ emission is partly clumpy and partly diffuse
on scales larger than about 1kpc. 
\oiii\ emission
is also detected at high significance, offset relative to the optical counterpart in the same direction as
the \cii\ clumps, but mostly not overlapping with the bulk of the \cii\ emission. The offset between different emission
components (optical/UV and different far-IR tracers) is similar to that which is observed in much more powerful starbursts
at high redshift.
We show that the \oiii\ emitting clump cannot be explained in terms of diffuse gas excited by
the UV radiation emitted by the optical galaxy, but it requires excitation by in-situ (slightly
dust obscured) star formation,
at a rate of about 7 \sfr. Within 20 kpc from the optical galaxy the ALMA data reveal two additional \oiii\ emitting
systems, which must be star forming companions. 
We discuss that the complex properties revealed by ALMA in the z$\sim$7.1 galaxy are consistent
with expectations by
recent models and cosmological simulations, in which differential dust extinction, differential excitation and 
different metal enrichment levels, associated with different subsystems assembling a galaxy, are responsible for
the various appearance of the system when observed with distinct tracers.}

% These dates will be filled out by the publisher
%\date{Accepted XXX. Received YYY; in original form ZZZ}

\authorrunning{Carniani et al.}
\titlerunning{Extended ionized and clumpy gas in a normal galaxy at z=7.1 revealed by ALMA}
   \maketitle
%%%%%%%%%%%%%%%%%%%%%%%%%%%%%%%%%%%%%%%%%%%%%%%%%%

%%%%%%%%%%%%%%%%% BODY OF PAPER %%%%%%%%%%%%%%%%%%

\section{Introduction}

Characterising the primeval galaxies of the Universe entails the  challenging goal of observing galaxies with modest star formation rates \citep[$<10$\sfr;][]{Salvaterra:2011, Finkelstein:2012, Dayal:2014, Robertson:2015} and approaching the beginning of the reionisation epoch ($z>6$). 
\\
To date, a large number of primeval galaxies have been identified thanks to the Lyman
Break technique \citep[e.g.][]{McLure:2013, Bradley:2014, Finkelstein:2015, Schmidt:2014, Bouwens:2015, Bouwens:2016, Oesch:2014, Oesch:2015a, Zitrin:2014, Calvi:2016, Infante:2015, Ishigaki:2015, Kawamata:2016, McLeod:2015}.
 To further our understanding on the formation and evolution of such primeval objects, we must investigate their nature and physical properties through multi-band
spectroscopic observations.

In the past few years, millimetre and submillimetre spectroscopy has proven
to be a powerful tool for studying the  interstellar medium and kinematics
of primeval galaxies  through the far-infrared (FIR) atomic fine structure lines, such as  \cii, [N{\sc ii}], \oiii, redshifted to millimetre wavelengths at $z>4$.
In particular, the \cii\ at 158$\mu$m lines is generally the strongest FIR emission line, tracing photodissociation regions \citep[PDRs;][]{Madden:1997, Kaufman:1999, Gracia-Carpio:2011, Pineda:2014}, neutral diffuse and partially ionised gas. 
This line has been detected in several quasars (QSOs), starburst, and submillimetre
galaxies (SMGs) at $z>4$ \citep{Maiolino:2005,Maiolino:2009, De-Breuck:2014, Maiolino:2012, Wagg:2012, Gallerani:2012,
Venemans:2012, Carilli:2013a, Carniani:2013, Williams:2014, Riechers:2014, Walter:2009, Cicone:2015}.
The high sensitivity of ALMA has recently enabled the 
detection of this FIR line also in faint and primeval star-forming galaxies
with SFR closer to the `normal' population (SFR$<100$ \sfr).
The properties of such galaxies in terms of \cii\ emission are heterogeneous.
A number of normal Lyman Break galaxies  at $5<z<6$   have been detected in \cii\
and it has been claimed that they follow the same \cii--SFR relation of  local galaxies \citep{Willott:2015, Capak:2015}.
However, several 
\cii\ detections or upper limits
imply the existence of galaxies at $z\gtrsim6$ well below the \cii-SFR relation 
\citep{Ota:2014, Schaerer:2015a, Gonzalez-Lopez:2014,
Maiolino:2015, Pentericci:2016, Knudsen:2016}. For other galaxies the \cii\ detection is above
the local \cii --SFR relation, including \cii\ systems without
optical or near-IR counterpart \citep{Maiolino:2015, Aravena:2016a, Capak:2015, Knudsen:2016}. To make
the scenario even more complex, for a significant number of galaxies at z$\sim$6--7
the \cii\ emission appears to be significantly offset both in spatial position and in
velocity.
A key result in this context has been our ALMA detection of \cii\ emission
associated with the normal star forming galaxy (SFR$\approx$6 \sfr)
\bdf\ at z$=$7.1 \citep{Castellano:2010, Vanzella:2011}.
This galaxy shows \cii\ emission fully consistent with the \lya\ redshift, but spatially
offset by 4 kpc relative to the optical emission \citep{Maiolino:2015}.
%The observation was explained in terms of a combination of stellar feedback and/or
%low gas metallicity of the optical star forming galaxy, while the offset \cii\ clump
%was explained in terms of a satellite, likely accreting, gas clump, as expected by some
%models \citep{Vallini:2015}.

A promising alternative way to study the ISM properties in high-$z$ systems is given
by the   FIR line of oxygen \oiii\ at 88$\mu$m.
With an excitation temperature of 164 K and a critical density of about 510 cm$^{-3}$, the
\oiii\ is a good tracer of ionised gas in moderate density HII regions.
%In addition, since an ionising energy of 35.12 eV is required to create O++ from singly ionised oxygen, \oiii\ transitions mainly arises in HII regions.  
%
FIR infrared observations in the local Universe up to $z\sim0.05$ have revealed that the
\oiii\ to FIR continuum luminosity ratio ranges between 0.03\% and 2\% \citep{Malhotra:2001, Negishi:2001, Brauher:2008} and in low-metallicity galaxies ($Z<1/3 Z_\odot$) the oxygen line can be up $\simeq$3 times brighter than that of \cii\ \citep{Cormier:2012}. 
Recent models have highlighted that \oiii\ is a key diagnostic of low metallicity ISM, especially in primeval galaxies characterised by very young stellar populations \citep{Inoue:2014, Vallini:2016}. 
In the distant Universe, the \oiii\ line has been  detected in two gravitationally lensed
AGN/starburst composite systems at redshift $z = 2.8$ and $z=3.9$ \citep{Ferkinhoff:2010}.
The only one \oiii\ detection at $z>4$ has been reported by \cite{Inoue:2016}
who have detected the \oiii\  with ALMA in a star-forming galaxy at $z\sim7.2$ and from which
they estimate an oxygen abundance of about one-tenth Solar.

We have obtained new ALMA \cii\ and \oiii\
observations of the star-forming galaxy \bdf\  at $z=7.1$. This galaxy
is located in a region characterised by a significant overdensity of
LBGs at z$\sim$7 \citep{Castellano:2016a}. Assuming no dust extinction,
this galaxy has a SFR=5.7~\sfr \citep{Maiolino:2015} and 
previous ALMA observations have revealed a displaced \cii\ emission relative to optical
emission. The offset is not ascribed to astrometric uncertainties, as astrometry
is checked through serendipitous sources.
The new ALMA \cii\ observations have a higher angular resolution and, along with
the \oiii\ data, enable us
to investigate more in detail the properties of the interstellar and
circumgalactic medium of this primeval system.

The paper is organised as follow. Section~\ref{sec:observations}  describes the ALMA observations. In Section~\ref{sec:continuum} we analyse the continuum emission at 410 GHz and 230 GHz. 
We investigate the properties of \oiii\ and \cii\ emission in the Section~\ref{sec:oiii} and \ref{sec:cii}, respectively. 
In Section~\ref{sec:discussion}, we interpretate our results and we summarise the study and give our conclusions in Section~\ref{sec:summary}.

Throughout this work we adopt the following cosmological parameter 
${\rm H_0 = 67.3 \  km s^{-1} \  Mpc^{-1} , \ \Omega_M = 0.315,  \ \Omega_{\Lambda} = 0.685}$ \citep{Planck:2014}.

\section{Observations}\label{sec:observations}

ALMA observations were obtained in Cycle 2 within the project  2013.1.00433.S. 
During this ALMA project \bdf\ was observed both in Band 6 (230 GHz) and in  Band 8 (410 GHz).

In Band 6 we re-observed the \cii\ emission in \bdf\ aiming at achieving the same sensitivity of previous observations obtained in Cycle 1 \citep{Maiolino:2015, Carniani:2015}, but with higher ($\times 8$) spatial resolution.  
ALMA Band-6 observations were carried out on 2015 June 30 and July 1 with a extended configuration (longest baseline = 1,570 m) and a  precipitable water vapour of PWV = 0.8 mm. 
The on-source integration time was about 3 hours.
The observations were performed in frequency division mode (FDM) with a total bandwidth of 7.5 GHz, divided in to four spectral windows (SPWs) of 1.875 GHz.
One of the SPWs was centred on the \cii\ redshifted frequency, which was observed with a spectral resolution of 1.95 MHZ ($\sim$2.5 km/s).

In addition to the Band 6 observations, the ALMA project aimed at mapping the \oiii\ at 88$\mu$m, which is redshifted to  418.5 GHz  at $z=7.107$.
Band 8 observations were obtained on 2015 May 1 and 15. 
The array configuration was composed by  55 12-m antennas with baseline length  in a range between 15 m and 560 m. 
The total on-source observing time was about  73 minutes and
the PWV was 0.5-0.6 mm.
The observations were carried out in FDM and the spectral band was set up to a spectral resolution of 1.95 MHz ($\sim 1.4$ \kms).
One  of the four SPWs was centred on the expected redshifted frequency of the \oiii.

The bandpass calibrator was J2258-2758 for both ALMA bands.
Neptune and Ceres were observed as flux calibrators, and J1924-2914 and J2223-3137 observation were used for phase and gain calibration in the Band 8 and 6, respectively. 
Unfortunately,  the phase calibrator J2223-3137 observed in Band-6 suffered from
instrumental errors (differential timing among the different antennae, resulting
in phase shifts across the field of view), therefore it has been not suitable to calibrate and correct the variation of phase and amplitude with time.
We therefore used the bandpass calibrator  J2258-2758, which was observed regularly during the ALMA programme, as phase calibrator in Band 6.  
Since the bandpass calibrator is usually observed for time scales shorter  than those of phase calibrator, the final phase errors are
larger than the expected ones and  the  sensitivity of  final images is worse than that requested. 
Indeed, the main effects of phase errors on interferometry are the loss of sensitivity due to decorrelation and the degradation of image quality due to scattering of flux throughout the image.
In conclusion,  the Cycle 2 Band-6 high resolution data have a sensitivity lower than that obtained in Cycle 1, although the on-source time and the
weather conditions are  similar.

Band 8 and 6 data were calibrated by using the {\sc casa} software version v4.5.3. 
We produced  final continuum images and cubes with the routine {\sc clean}, selecting a natural weighting to optimise the sensitivity. 
Because a bright serendipitous source \citep[flux density S$_{\rm Band-6} = 400$ \ujy;][]{Maiolino:2015,Carniani:2015} is present in the \bdf\ field of view, the cleaning of the continuum maps was performed with 500 iterations.
No cleaning was performed in any cube, since there are no bright line emission in the field
of view and the continuum flux of the serendipitous sources is not high enough
to enable channel-by-channel cleaning.
The achieved sensitivities and angular resolution  are listed in the Table~\ref{tab:ALMA}.
The sensitivity is about a factor 1.5 worse than achieved in Cycle 1 at lower angular
resolution.

Finally, we combined Band-6 data obtained in Cycle 1 to the new one observed in Cycle 2. 
Before combining the visibilities of two observations, we manually re-scaled their relative weights so that they matched the noise variance measured in the two datasets.
The combined data set results into a continuum sensitivity of 7 \ujy\ and a combined
angular resolution of about 0.6\arcsec.  

In this work we also make use of the Hubble Space Telescope (HST) F105W (Y-band hereafter) imaging of the BDF field obtained under cycle 22 programme 13688 and described in \cite{Castellano:2016a}. The relevant S/N=10  magnitude limit is  Y$_{\rm AB}$=28 in apertures of 0.38\arcsec,  which is two times larger than the HST angular resolution in Y-band (0.19\arcsec).
Following \cite{Maiolino:2015} we matched the ALMA observations to the HST images and we
applied a position shift to the Y-band by $\sim0.4\arcsec$ toward the South-East, based on the consistent
continuum
offset measured on the two serendipitous sources.

\begin{table*}
\caption{Summary of the ALMA observations}           
\label{tab:ALMA}      
\centering          
\begin{tabular}{l c c c c c c}    
\\
\hline
Observations & Frequency  & $\sigma_{\rm cont}$ &  $\sigma_{\rm 100km/s}$ & Beam & angular & Primary\\
  & [GHz]  &  [\ujy/beam] & [\ujy/beam] & & resolution  & Beam\\
\hline

Band 8 Cycle 2   & 410 & 40 & 260 & $0.5\arcsec\times0.4\arcsec$ &    & 16\arcsec \\
\hline
Band 6 Cycle 1   & 230  & 8  & 62 & $0.8\arcsec\times0.6\arcsec$ & low & 22\arcsec \\
Band 6 Cycle 2   & 230  & 11 & 90 & $0.3\arcsec\times0.2\arcsec$ & high & 22\arcsec\\
Band 6 Cycle 1+2 & 230 & 7  & 58 & $0.6\arcsec\times0.5\arcsec$  & medium & 22\arcsec\\

\hline
\\
                
\end{tabular}   
\end{table*}

\section{Continuum emission}
\label{sec:continuum}

\begin{figure}
\centering
 \includegraphics[width=0.8\columnwidth]{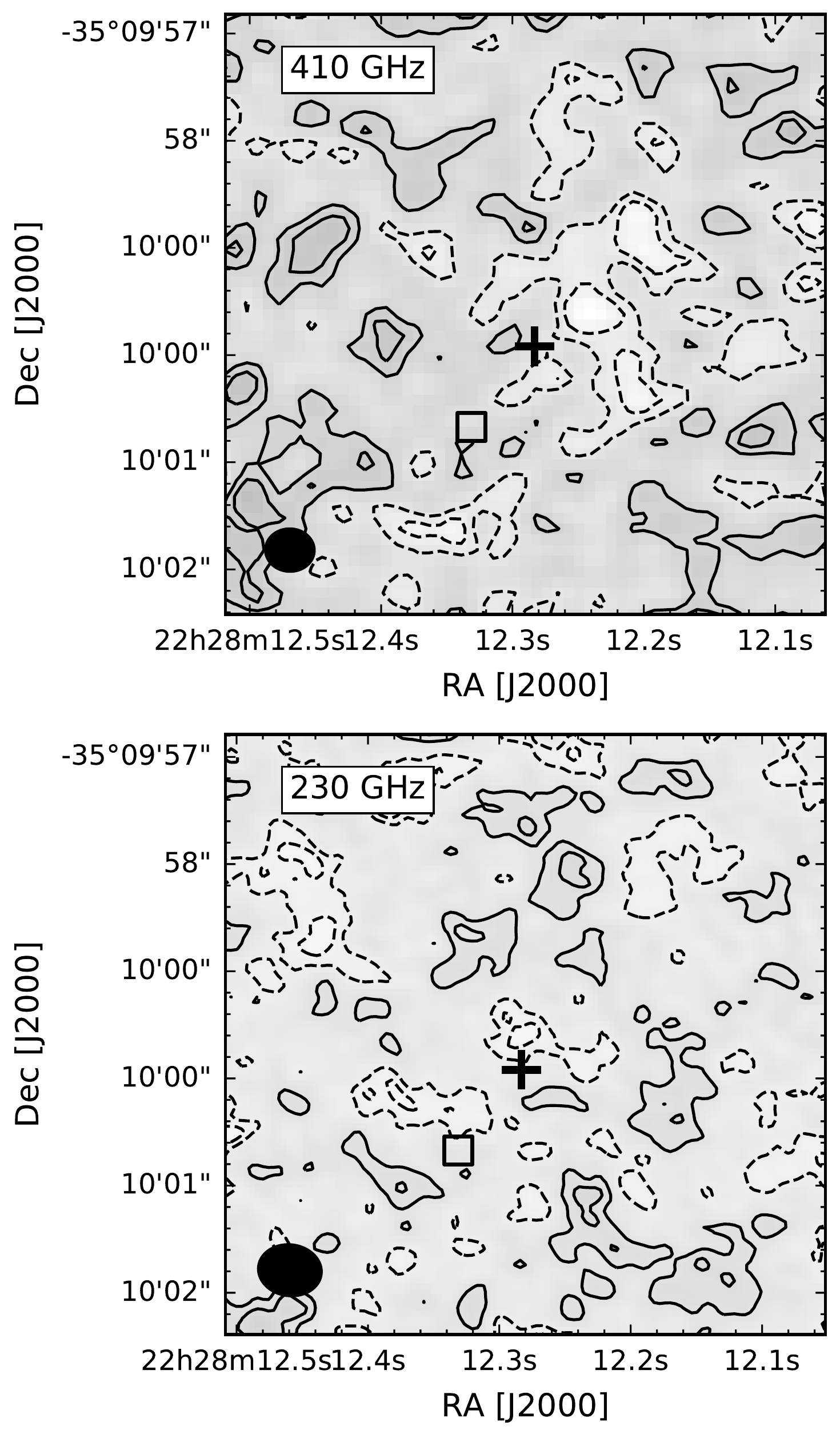}
 \caption{ALMA continuum map in band 8 (410 GHz; top) and in band 6 (230 GHz, bottom). Black solid contours are at levels of 1 and 2$\sigma$, where
 i.e $\sigma=40$ \ujy/beam and $\sigma= 7$ \ujy/beam in the two bands, respectively. Dashed contours trace negative levels at --1 and --2$\sigma$.
The synthesised beams are shown in the bottom left corners. The position of \bdf\ measured in the HST Y-band is marked with a cross. The square mark shows the location of the \cii\ detection.
 }
 \label{fig:continuum}
\end{figure}

Figure~\ref{fig:continuum} shows the ALMA continuum maps at 410 GHz (Band 8) and at 230 GHz (Band 6) around the target \bdf.
Band-6 image is obtained combining Cycle 1  to Cycle 2 data (sensitivity and angular resolution are listed in Table~\ref{tab:ALMA}).
The FIR emission is not detected in either continuum maps at the location of \bdf.
%, while the two brightest serendipitous sources detected by \cite{Carniani:2015} are visibile with a S/N$>3$.

At this rest-frame wavelengths the emission is due to dust thermal emission heated by the radiation field of young
stars. The
non detections indicate that \bdf\ is characterised by low dust
content ($M_{\rm dust}<5\times10^{7}$), as already discussed by \cite{Maiolino:2015}.
A similar result has been  claimed in recent studies of  high-$z$ ($>6$) star-forming galaxies, in which the FIR emission relative to the
UV is similar to that observed in nearby dwarf or irregular galaxies \citep{Walter:2012, Kanekar:2013, Ouchi:2013, Gonzalez-Lopez:2014, Ota:2014, Schaerer:2015a, Willott:2015, Bouwens:2016}.
%The lack of dust can be associated to the lower metallicity.
 
Assuming a dust temperature range between 25 K and 45 K \citep{Schaerer:2015a}, we derive a 2$\sigma$ upper limit estimates for the
infrared luminosity $L_{\rm FIR}<10^{11}$ \lsun. If we adopt the ``standard'' FIR--SFR conversion factor for local galaxies
\citep{Kennicutt:2012}
this upper limit on the FIR luminosity translates into an  2$\sigma$ upper limit on the IR-derived star formation rate of SFR$<$12 \sfr,  which is consistent with the SFR inferred for the rest-frame UV continuum (5.7 \sfr). However,
since the galaxy
is dust-poor, the upper limit on the FIR luminosity likely translates into a higher limit on the SFR.

\section{\cii\ emission} 
\label{sec:cii}

In this section we compare the  maps of the carbon line obtained with different ALMA
configurations: semi-compact array \citep[low angular resolution;][]{Maiolino:2015},
extended array (high angular resolution) and combined data. We recall  that the
previous \cii\ observation of this source had an angular resolution of about 0.7$''$,
while the sensitivity was higher than the new high resolution data, as reported in Table~\ref{tab:ALMA}.

\begin{figure}
 \quad \quad \quad \includegraphics[width=0.75\columnwidth]{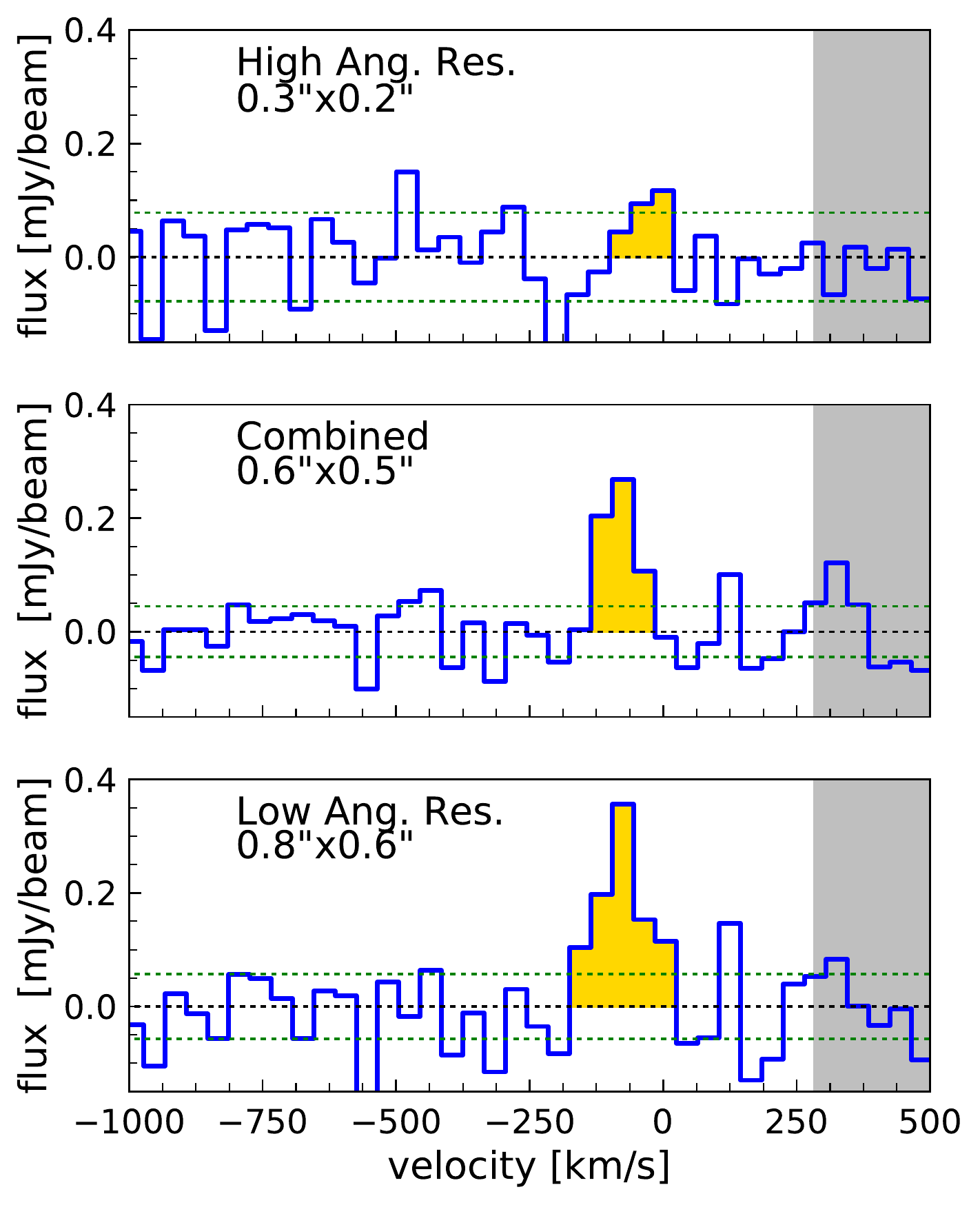}\\
\Large  (a)\\ \\
  \includegraphics[width=0.9\columnwidth]{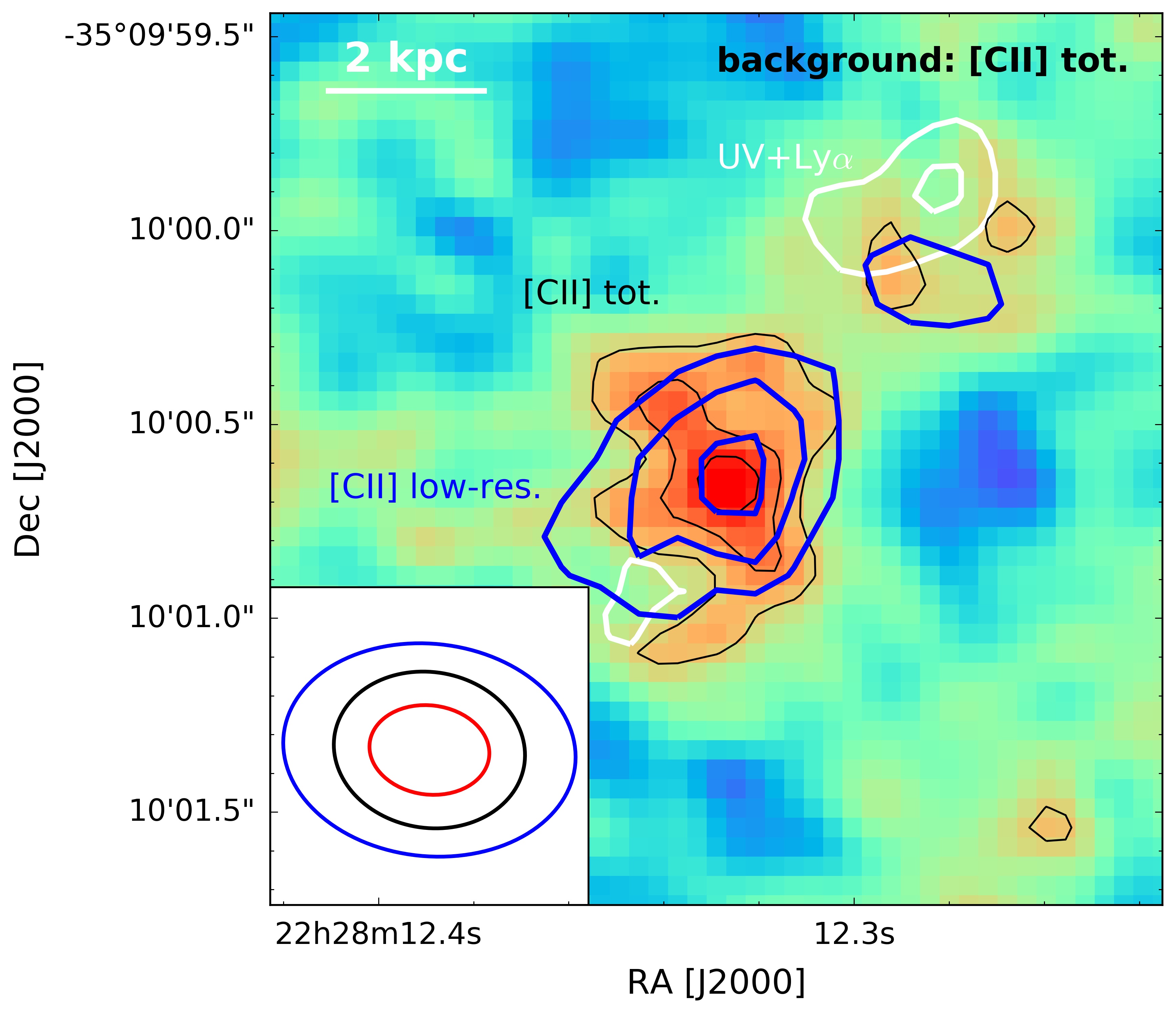}\\
 \Large (b)

\caption{ (a): ALMA spectrum at three different angular resolution (high, medium and low) extracted at the  peak position of the \cii\ emission  with a spectral resolution of 40 km/s. 
 The rms noise levels in each spectrum is shown by the green dotted line.
 The grey shaded  region indicates the part of the spectrum affected by higher noise because of atmospheric absorption. (b): Map of the \cii\ emission obtained by
 combing the two datasets (medium angular resolution)  in which black contours are  in steps of 1$\sigma$ = 5.6 mJy/beam km/s, starting at 2$\sigma$. Blue
 contours trace the \cii\ surface brightness using the low angular resolution data only \citep{Maiolino:2015}  and contours are at levels 2, 3 and 4 times noise per beam.
The white contours trace the Y-band emission (UV-rest frame and Ly$\alpha$). 
 The bottom-left corner shows the synthesised beam of the high (red), medium (black) and low (blue) angular-resolution datasets, respectively (see Table~\ref{tab:ALMA}).}
 \label{fig:cii}
\end{figure}

 We compare the spectra  at the position of the \cii\ detections obtained from different ALMA array configurations.
The top panel of Figure~\ref{fig:cii}a shows the spectrum of the \cii\ emission extracted from the  dataset with high angular resolution ($\sim$0.3\arcsec$\times$0.2\arcsec). 
The \cii\  is  only marginally detected (S/N$\sim$2). This is only partly due
to the lower sensitivity of the new observations. The comparison with
the spectrum obtained with lower resolution observation obtained in Cycle 1
(bottom panel of Fig.~\ref{fig:cii}a)
confirms that the high resolution data also miss a significant  fraction of
the line flux, which is an indication that the total emission is not powered by a compact source with size smaller than 0.3\arcsec ($\sim$1.6 kpc). The diffuse \cii\ emission is resolved out in the ALMA extended  configuration (see Appendix~\ref{sec:appa}).
%By measuring the line flux (Table~\ref{tab:cii}) we derive that  about  70\%  of the emission is resolved out in the 
%ALMA extended configuration. 
By measuring the line flux (Table~\ref{tab:cii}) we derive that  most of the  total emission comes from a more extend region.
We conclude that about  70\% ($\pm$20\%) of the flux in the
\cii\ system is extended and diffuse on scales larger than about 1 kpc, which is consistent  with the idea that the bulk  of the \cii\ emission is associated
 with relatively diffuse gas on scale larger than the star formign regions, 
 as expected by recent model of primeval galaxies
 \citep[e.g.][]{Vallini:2013, Vallini:2015, Pallottini:2016, Katz:2016}.

By combining the visibilities of the two observations obtained in Cycle 1 and 2 we generate a single
cube  whose resulting angular resolution is $\sim0.6\arcsec$ and a  sensitivity of 58 \ujy/beam in spectral
bins of 100km/s (Table 1),
 which is slightly higher than the sensitivity reached in the two separate datasets. 
The \cii\ spectrum obtained from the combined data is shown in the second panel of Figure~\ref{fig:cii}a.
 By integrating the line under the shaded gold region, we obtain a significance of the detection of 6.5\footnote{This is the ``spectral''
significance, i.e. relative to the noise in the same spectrum.}.
The flux is extracted in the same aperture as the spectrum in \cite{Maiolino:2015} and we applied
the aperture correction estimated by using the bright serendipitous source as in \cite{Maiolino:2015}.
We note that the integrated flux  obtained from the combined observations (Table~2) is  consistent  with the estimates obtained by the low angular resolution data alone previously reported
by \cite{Maiolino:2015}. The [CII] line in the combined dataset is slightly narrower (FWHM = $75\pm$20 km/s) than previously
reported in the lower resolution data (FWHM = $102\pm$21 km/s), but still consistent within uncertainties.
The map resulting from the combined set is shown in Figure~\ref{fig:cii}b,  which, as expected, is
clearly characterised by a more {\it clumpy} structure
than the low-resolution map (shown in blue). It should be noted that the signal to noise of the
peak emission  (4.8$\sigma$, Table~2) is not as high as the integrated flux significance,
simply because the emission is resolved and extended, hence the significance of the line detection
is higher than what inferred by the flux/beam of the emission peak in the map. 
Additionally, the significance of the peak emission is slightly higher than what previously found in the lower
resolution data alone (Table 2), which can be explained in terms of beam dilution of the latter and
higher sensitivity to clumpy emission of the former. 

 Comparing the map and the spectra obtained with different ALMA array configuration, we conclude that the new observations confirm that the spatially-offset \cii\ emission is spatially extended and reveal that such extension is due both to multiple smaller ($\lesssim1-2$ kpc) clumps, distributed on such large scales, and to diffuse extended emission. 

We finally note that at the location of the optical (Y-band) counterpart the new combined map confirms
the marginal detection ($\sim 2\sigma$) of some \cii\ emission.

\begin{table}
\caption{Properties of the \cii\ emission in \bdf\ for the different datasets obtained with ALMA.}     
\tiny      
\label{tab:cii}      
%\centering          
\begin{tabular}{l c c c }    
\\
\hline
&  Low AR  & High AR &Combined \\
&  Dataset  & Dataset  & Dataset \\

\hline
\\
%$\nu_{\rm obs}$ [GHz] & 234.43 & \\
$\Delta V$ [km/s] & $-71\pm10$ & - & $-64\pm10$\\
FWHM [km/s] & $ 102\pm21$ & - & $75\pm20$ \\
Peak emission [mJy/beam km/s] & 29.1$\pm6.4$  & $3.0\pm1.3$ &  26.8$\pm$5.6\\
Flux$^\star$ [mJy km/s] &  $49\pm7$ & $15\pm8$ & $39\pm6$\\
Luminosity [10$^{7}$ \lsun] & $5.9\pm0.8$ & $1.9\pm1.0$ & $4.9\pm0.6$ \\

\\
\hline
                
\end{tabular}
\tablefoot{$^\star$ Following \cite{Maiolino:2015} we have applied a flux  aperture correction since the extraction aperture has a size close to the beam. }
\end{table}

\section{\oiii\ emission}
\label{sec:oiii}

\subsection{Line detection}
We have searched for \oiii\ emission in the band 8 cube, within a few 100~km/s of the Ly$\alpha$
emission.
We detect three \oiii\ emission line systems. On the flux map (in units of flux/beam) their emission peaks have a
signal-to-noise ratio S/N$\goa5$.
The integrated 
line emission is detected at a level of confidence  $>6.5~ \sigma$. The reliability of these detections,
including the comparison with the ``negatives'' detection rate, is discussed in the Appendix.

\begin{figure*}
 \includegraphics[width=2\columnwidth]{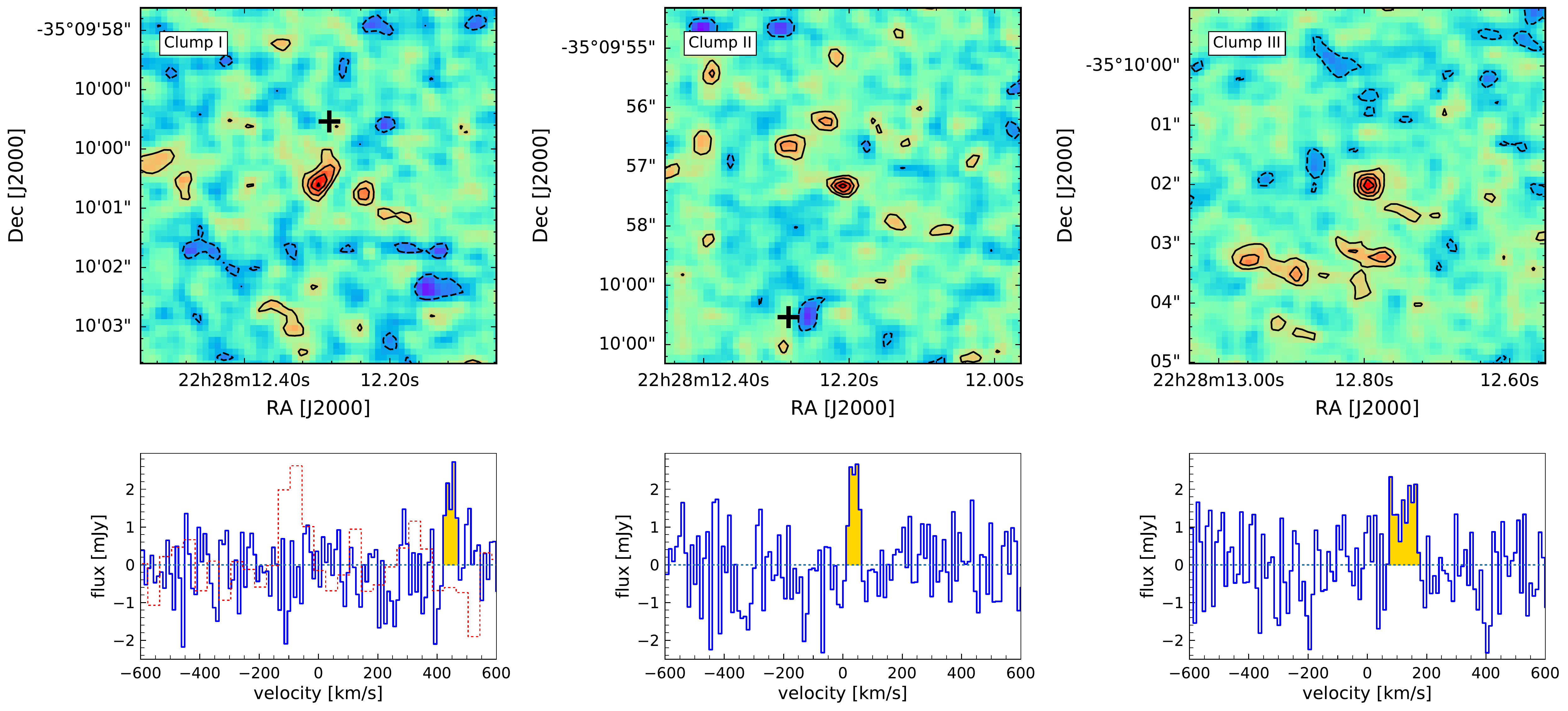}
 \caption{ 
The top panels show the flux maps of the three \oiii\ detections, clump-I, -II,
and -III (from left to right). Black contours show emission at levels of 2,3,4
and 5 $\sigma$. The location of the \bdf\ galaxy is marked as a black cross (in the third panel the mark is outside of the field of view).
The bottom panels show the spectra of the line emitters at a spectral resolution of 10 km/s. The yellow shaded region shows the region used to
integrate the flux to obtain the maps in the upper panels. Velocities are relative to the \lya\ peak of \bdf.
 The red dotted spectrum in the left panel show the [CII] spectrum shown in Figure~\ref{fig:cii} and scaled  arbitrarily up in flux density for comparison. 
}
 \label{fig:oiii}
\end{figure*}

The properties of the \oiii\ detections are summarised in Table \ref{tab:oiii} and the 
\oiii\ flux maps of the three detections are shown in the top panels of Figure~\ref{fig:oiii}, obtained by
integrating the line profile under the velocity range highlighted by the shaded region in the respective spectra (bottom panels of Figure~\ref{fig:oiii}). 
The black cross in the maps shows the location of the optical counterpart of
\bdf\ from the HST Y-band image.  We note that the significance of the peak emission in the maps (third column
of Table \ref{tab:oiii}) is slightly lower then
the line integrated significance (second column of Table \ref{tab:oiii}) because the emission is spatially slightly resolved.

The \oiii\ clumps are located within $\sim6.5\arcsec$ ($\sim34$ kpc) from the star-forming galaxy \bdf\ and have velocities within 500 km/s relative to the \lya\ peak (Table~\ref{tab:oiii}).
One of the three detections (clump-I in
Table~\ref{tab:oiii})
is located very close to the optical Y-band counterpart of BDF-3299, but it has a small offset
of 0.6\arcsec\ (i.e. 3.5~kpc) in the same direction as the \cii\ offset. However, as shown in Fig.4, it is not
coincident with \cii\ either.
The \oiii\ emission is only marginally overlapping with the \cii\ emission, the two emission peaks being offset
by 0.2\arcsec \ ($\sim$1~kpc), which is slightly larger than the  uncertainties of 0.15\arcsec\ based on the significance of the detections and beam size.
 
\begin{table*}
\caption{Properties of the \oiii\ clumps observed in the field of view of \bdf.}           
\label{tab:oiii}      
\centering          
\begin{tabular}{l c c c c c c c }    
\\
\hline
Clump & S/N$_{integrated}^{\rm(a)}$ & S/N$_{peak}^{\rm (b)}$ &  FWHM & $v_{\rm [O\textsc{iii}]}-v_{\rm Ly\alpha}$ & F(\oiii) & L(\oiii) & Distance \\  
&  &  & [\kms] & [\kms] & [mJy \kms] & [$10^8$ \lsun] & [\arc]\\ 
\hline
\\
%flux estimates are beam corrected
clump-I    & 7.2 &  5.0 & $40\pm10$  &  $440\pm5$  & $85\pm12$ & $1.8\pm0.2$   & 0.7  \\
clump-II    & 6.9 &  5.3 & $40\pm10$  &  $32\pm5$  & $100\pm14$ & $2.2\pm0.3$  & 2.5\\
clump-III    & 6.7  &  5.6 & $80\pm10$  &  $110\pm10$  & $260\pm40$ & $5.8\pm0.6$  & 6.5\\
\\
\hline               
\end{tabular}
\\
\tablefoot{{\footnotesize $^{\rm(a)}$ The integrated significance is the ratio between the integrated flux of the line
and the relative error calculated in the extracted
spectrum. $^{\rm(b)}$ Significance of the line emission ``peak'' in the collapsed image. The peak emission significance is slightly lower than
the line integrated significance because the emission is spatially slightly extended.} }
\end{table*}

\begin{figure}
 \includegraphics[width=\columnwidth]{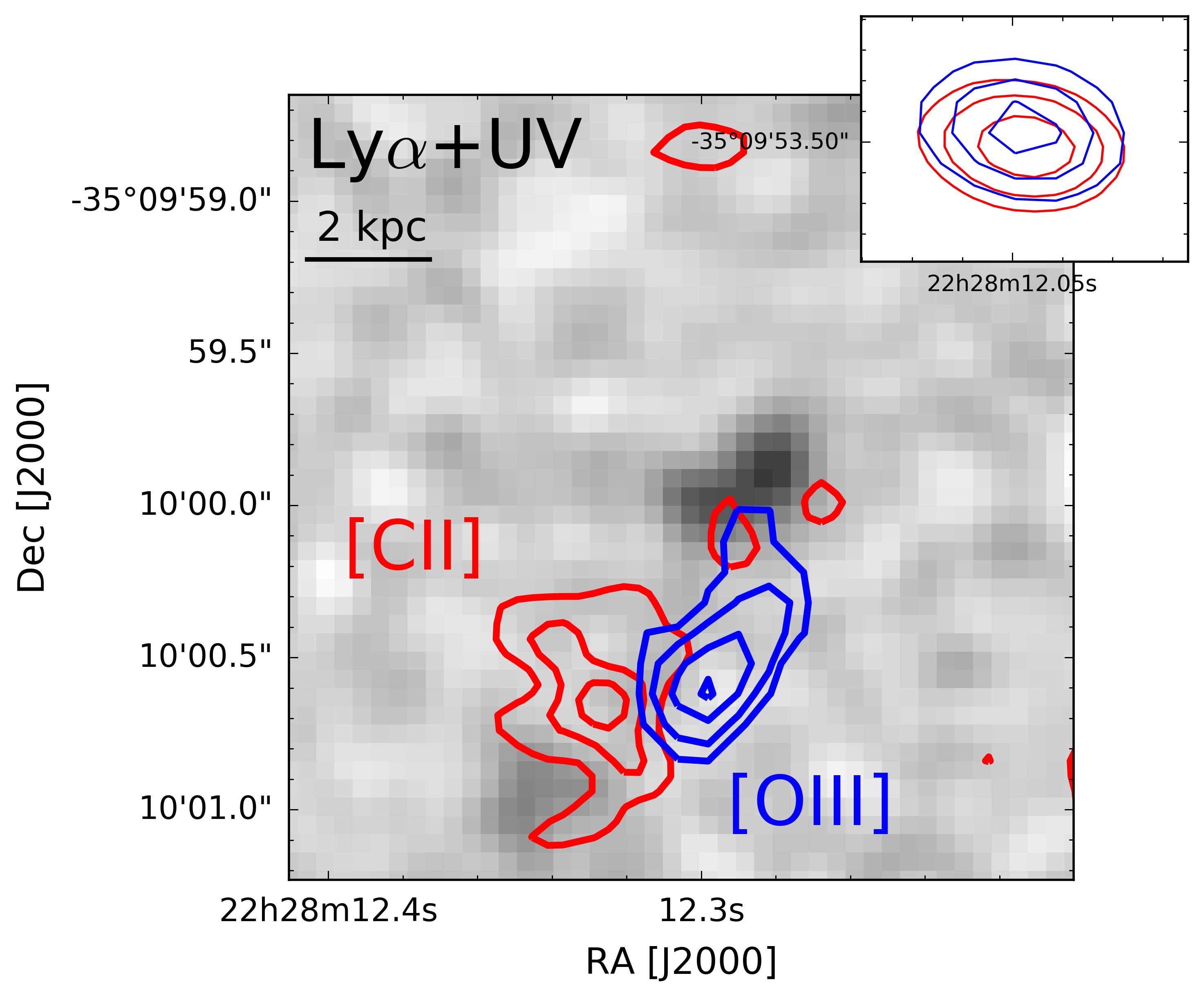}
\caption{ The HST Y-band image of the \bdf\ field is shown in the background in grey scale.
%The black cross indicates the location of the galaxy \bdf\ at z=7.109. We recall that at this redshift the Y-band samples both \lya\ and UV continuum. 
The blue contours show the \oiii\ map of clump-I. 
The red  contours show  the \cii\ emission. 
In both maps the
contours are at levels of 2, 3, and 4 $\sigma$. Since the emission is resolved in both cases, the peak emission is not representative of the
significance of the emission when extracted from the entire emission region. The inset shows in blue and red contours the continuum emission of the bright
serendipitous source at North-West in Band-6 and -8, respectively, showing that the astrometry in Band-8 is consistent within the error (0.1\arcsec) with the
astrometry in Band-6.
}
\label{fig:oiii_distribution}
\end{figure}

To verify that the positional offset between the  \cii\ and \oiii\ emission is not caused by an astrometric
issue, we compare the centroid of the  continuum emission in Band-6 and -8 of the  brightest serendipitous  source located at north-west. 
The source is detected in Band 8 at a level of 33$\sigma$ and its centroid position  is consistent with that
in Band 6 (inset in Figure~\ref{fig:oiii_distribution}). This nice agreement confirms that the spatial offset between the two clumps is real.
In addition, the central velocity of the \oiii\ is redshifted by about 440 \kms\ with respect to the \lya\ peak and about 500 \kms\ relative to the \cii\ emission.
The two emission lines, \oiii\ and \cii, thus are tracing two different clumps. 
 The far-infrared properties of the optical galaxy and the two clumps are summarised in the Table~\ref{tab:summary}.
The spatial and spectral offsets between \oiii, \cii\ and \lya\ emission will be discussed in details in Section~\ref{sec:discussion}.

\begin{table}
\caption{Summary of the far-Infrared properties at the location of the optical gaalxy,  \cii\ and \oiii\ clumps.}           
\label{tab:summary}      
\centering          
\begin{tabular}{l c c c }    
\\
\hline
 & optical clump & \cii\ clump & \oiii\ clump \\
S$_{\rm230GHz}$ {\tiny [$\mu$Jy]}& $<21$ & $<21$ & $<21$\\
S$_{\rm [C\textsc{ii}]}$ {\tiny [mJy km/s]} & $<19$ &  39$\pm$6 & $<$ 19\\
L$_{\rm [C\textsc{ii}]}$ {\tiny [$10^7$ L$_\odot$]} & $< 2$ &  $4.9\pm0.6$ & $< 2$\\
%$v_{\rm [C\textsc{ii}]}$ & 0 & -64 & \\
%FWHM$_{\rm [C\textsc{ii}]}$ & 100 & 75  &\\
S$_{\rm [O\textsc{iii}]}$ {\tiny [mJy km/s]} & $< 30$ & $< 30$  &  85$\pm$12 \\
L$_{\rm [O\textsc{iii}]}$ {\tiny [$10^8$ L$_\odot$]} & $<5.5$ & $<5.5$ &   1.8$\pm$0.2 \\
%$v_{\rm [O\textsc{iii}]}$ & 0 & & 440 \\
%FWHM$_{\rm [O\textsc{iii}]}$ & 100 & & 40\\
\\
\hline               
\end{tabular}
\\
\end{table}

\subsection{Reliability of \oiii\ detections} 

In this section we discuss the significance of the \oiii\ detections checking whether negative line emitters are detected with the same significance or not.

\begin{figure}
 \includegraphics[width=\columnwidth]{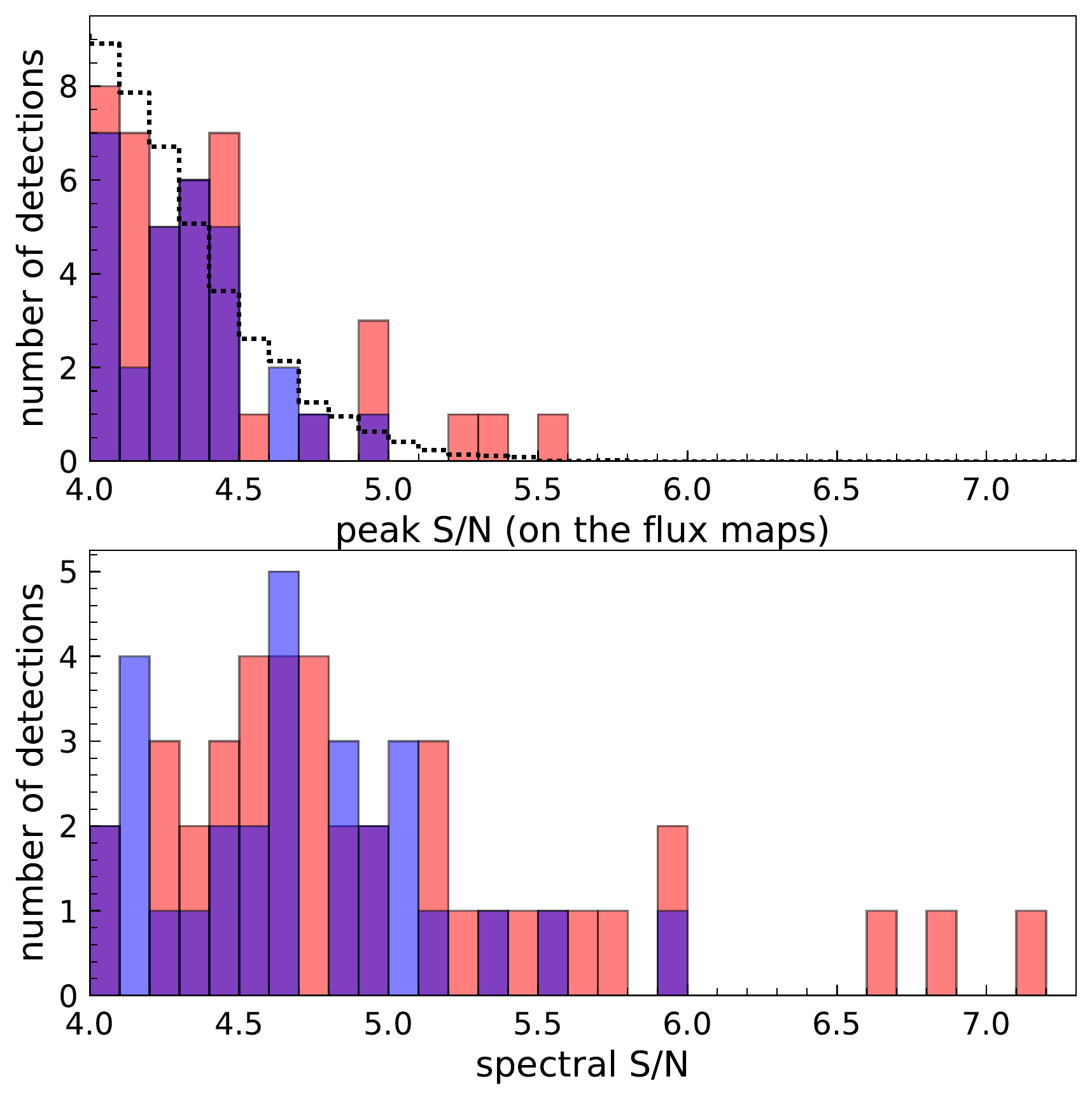}
 \caption{{\it Top}:  Number of positive (red) and negative (blue) detections as a function of S/N, which is defined as the ratio of the peak emission to the noise in the flux map.  The dotted black line shows the  distribution of positive (or negative) detections obtained from 100 pure noise datacubes and normalised to one datacube. {\it Bottom}:  Number of positive (red) and negative (blue) detections as a function of ``spectral'' significance, i.e. relative to the noise in the same spectrum, which is defined as the ratio of the flux line to the relative error.
 }
 \label{fig:pos_neg_distribution}
\end{figure}

We  performed a blind search for positive and negative line emitters within the primary beam area (FWHM = 16\arcsec) and within a velocity range  $\mid v\mid<$500 \kms\  relative to the redshift of the \lya\ peak. 
 We  excluded the region within 2\arcsec\ from the  continuum serendipitous source, since the uncleaned continuum map shows some low-level sidelobe residuals in this area.
In the defined velocity range, our blind line search algorithm scans the ALMA datacube rebinning it in different velocity channels,  from  30 \kms\ to 300 \kms, in step of 15 \kms. 
In each averaged velocity channel plans we computed the rms and we searched for positive (negative) peaks exceeding a S/N=4 (-4).
Top panel of Figure~\ref{fig:pos_neg_distribution} shows the distribution of positive (red) and negative (blue) detections as a function of S/N of
the ``peak'' in each map.
At S/N$\gtrsim$5, there are six potential positive detections and only one negative. Since the positive ones could be spurious because of noise fluctuations, we compare this distribution to that expected in blank fields. We simulated 100  blank datacubes similar to the real data and we applied on them the same source extraction algorithm used in the ALMA observations. In the simulated data, the distribution of positive peaks is similar to that of  negative ones and we expect to have only one negative and positive detections per datacube at S/N$\sim5$. This simulated noise distribution is shown with the dotted histogram in the top panel of Figure~\ref{fig:pos_neg_distribution}.  

We note that recent similar blind search works claim  a reliability level of 50\% at S/N=5 \citep{Walter:2016, Aravena:2016a}, so half of our detections at S/N$\geq$5 could be spurious. For that reason, we performed a further analysis.    
For each putative detections (positive and negative) we extracted the relative spectrum and we estimated the flux line and relative error in the extracted
spectrum.
We note that the negative line emitters have a line flux with a level of confidence below $6\sigma$ (bottom panel of Figure~\ref{fig:pos_neg_distribution}) and only three out of six positive detections with S/N$\gtrsim$5 in the previous analysis have level of confidence $>6$ as expected from the reliability level.
This support the reliability of the three selected putative positive detections $>6.5\sigma$.

\section{Discussion} 
\label{sec:discussion}

\subsection{Spatial and spectral offsets} 

The offset between \cii\ and \oiii\ emission, and also relative to the optical counterpart, may appear anomalous.
However, such offsets (either spatial, in velocity, or both) are  also seen in some high-z SMG galaxies
\citep{Riechers:2014, Decarli:2014, Pavesi:2016}.
 Recent ALMA observations have revealed
spatial and spectral offsets between  FIR-lines, such as \cii\ and \oiii, and  optical
emission \citep{Willott:2015, Capak:2015, Inoue:2016, Knudsen:2016} also in  normal
high-redshift galaxies ($z>5$), with modest SFR ($<100$ \sfr). Therefore, \bdf\ may not be a rare case.
These offsets have been only marginally discussed (or completely ignored) in the literature,
but they are probably revealing important physical properties of these high redshift systems.

In this section we make a brief statistical assessment of the offset between far-IR lines and
optical counterparts in normal galaxies at z$>$4 reported in the literature.

%In fact, some of these clumps may reveal star-formation in situ (see Section~\ref{sec:gas_mass}) that are not visible in the deepest HST near-IR images and only James Webb Space Telescope (JWST) will allow the identification of counterparts for these FIR-line emitters leading to an in-depth study on their properties.

\begin{figure*}
(a)
 \includegraphics[width=\columnwidth]{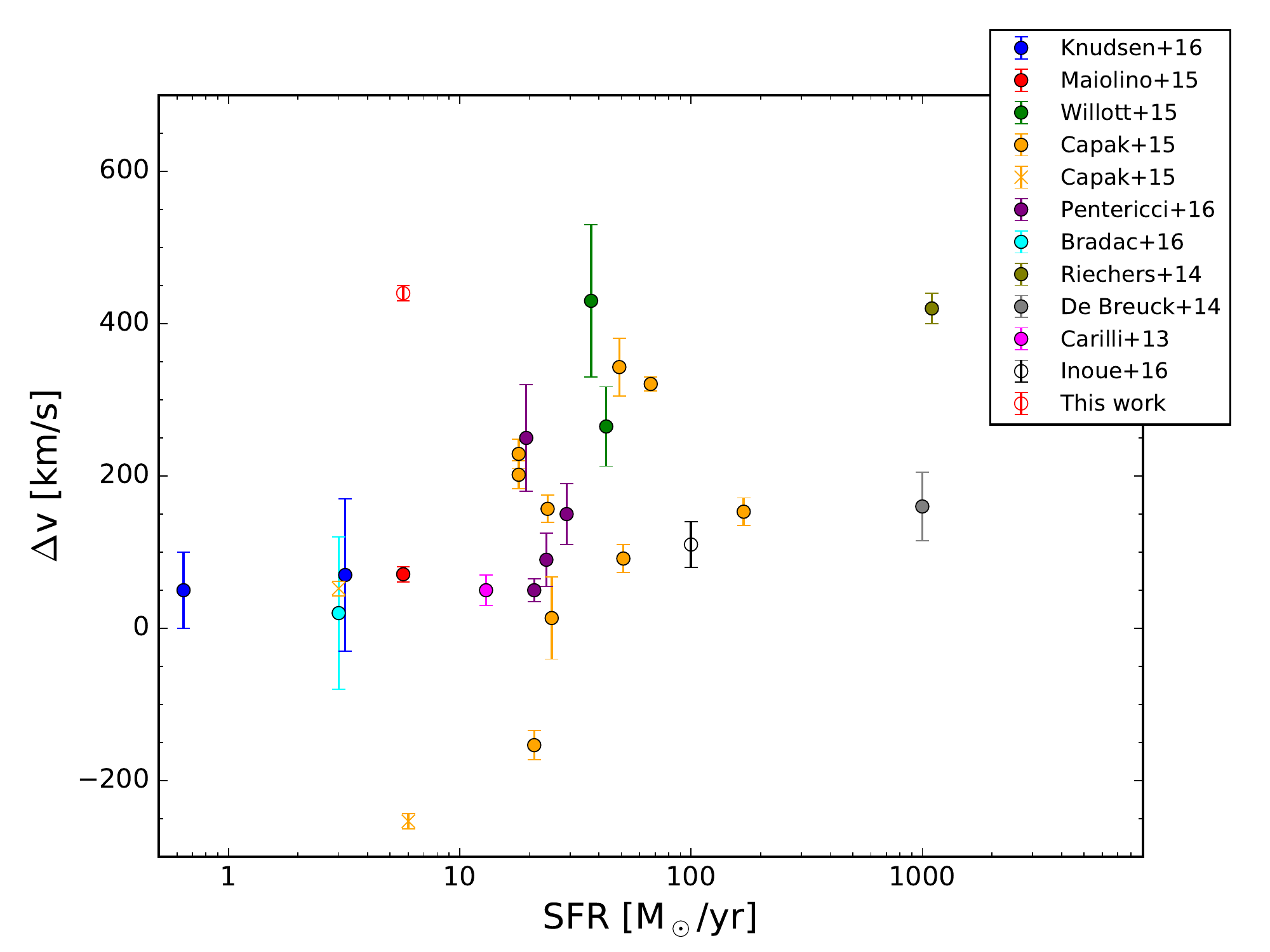} 
 (b)
  \includegraphics[width=\columnwidth]{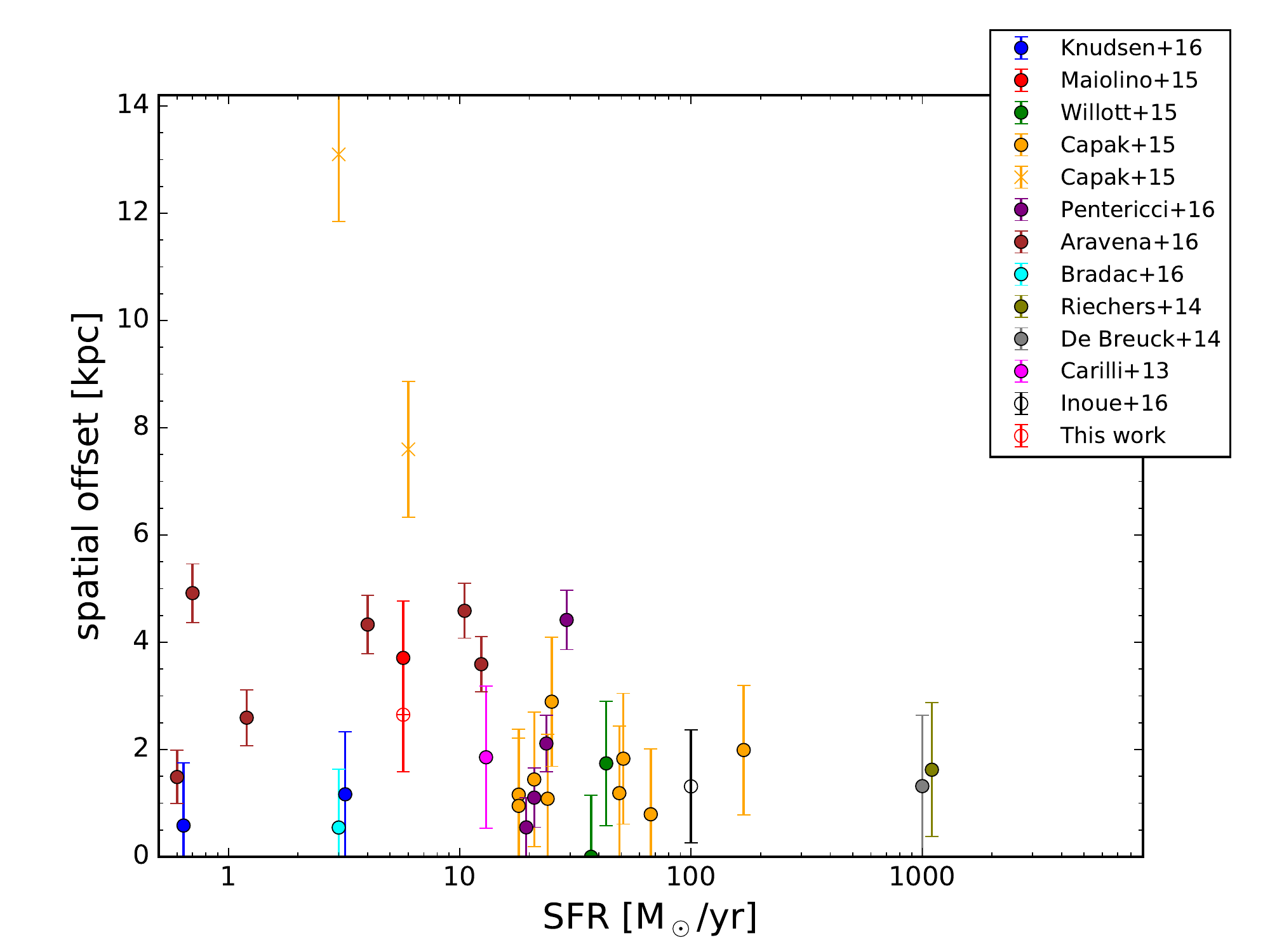} \\ 
%   (c)
%   \includegraphics[width=\columnwidth]{Figure/fig_3}
%   (d)
%    \includegraphics[width=\columnwidth]{Figure/fig_4}\\
 %   (e)
 %    \includegraphics[width=\columnwidth]{Figure/fig_5}
 %    (f)
 %     \includegraphics[width=\columnwidth]{Figure/fig_6} 
 \caption{ Spectral (a) and spatial (b) offsets between the far-IR fine structure lines and the optical/UV tracers in high redshift galaxies
 galaxies, from the literature.
 The filled circles are  estimated from \cii\ emission obtained from \protect\cite{Maiolino:2015, Willott:2015, Capak:2015, Knudsen:2016, Pentericci:2016, Aravena:2016a}, while the open ones denote the spectral and spatial offsets revealed from  \oiii\ detections \citep[this work and][]{Inoue:2016}. The orange cross symbols indicates the offsets of two sources \citep[HZ5a and HZ8W;][]{Capak:2015}  for which the association between \cii\ and the optical galaxy is uncertain.
 }
 \label{fig:offset}
\end{figure*}

In Figure~\ref{fig:offset} we summarise spectral and spatial offset from literature studies observing
SMG (SFR$>$1000 \sfr) $z>$4 \citep{Riechers:2014, De-Breuck:2014} and  primeval galaxies with SFR$<100$ \sfr and z$>$4 \citep{ Mallery:2012,Carilli:2013a, Williams:2014,  Maiolino:2015, Willott:2015, Capak:2015, Knudsen:2016, Knudsen:2016a, Pentericci:2016, Inoue:2016, Bradac:2016}.
Since this work does not focus on determining a correlation between the offsets and the galaxy properties, we also
report the positional offset revealed in in high-$z$ galaxies whose redshifts  have not been spectroscopically
confirmed in the optical yet \citep{Aravena:2016a}.
The Figure is composed of panels showing the spatial and spectral offsets  as a function of
SFR. 
Spatial and spectral offsets are present in most of these sources.
However no clear correlation is visible between offsets and galaxy properties.
It should be noted that, by 
analysing the first deep ALMA continuum image of the Hubble Ultra Deep Field (HUDF), \cite{Dunlop:2016} claim
that,
even after correcting for a systemic astrometric offset between the ALMA and HST astrometry,
the HST images have residual random positional errors with a $\sigma =0.2\arcsec$, which they claim can account
for offsets up to $\sim 0.5-0.6\arcsec$ (which is somewhat surprising given the small size of the area
and given that this is probably the deep field studied in greatest detail).
However, position offset larger than $\sim 0.5-0.6\arcsec$ cannot be associated to positional noise errors.
 In some studies the astrometry has been verified by using additional external datasets \citep[e.g.][]{Willott:2015}, while in our previous works we have verified the astrometry accuracy by  using serendipitous sources in the field of view. 
Finally, a number of
\cii-emitters are simply double (with one of the two sources having no optical counterparts), clearly indicating
that the absence of optical counterpart, cannot be ascribed to instrumental offsets \citep{Capak:2015}.
Large positional offsets between ALMA and HUDF optical and NIR observations have been also reported by \cite{Aravena:2016a}, who  found twelve \cii\ candidates with significance $>4.8\sigma$ located at a distance up to 1\arcsec\
from the optical counterpart.

In addition to positional offsets, Figure~\ref{fig:offset} indicates that there are velocity offsets
between the FIR lines and the optical/UV spectral features, spanning
a range between --300 km/s and +420 km/s. Velocity shifts between optical/UV lines, and in particular relative
to Ly$\alpha$, have been found and investigated by several authors
\citep[e.g.][]{Pettini:2002, Shapley:2003,
Gallerani:2016, Stark:2017,Williams:2014, Erb:2014}. 
In the case of Ly$\alpha$ IGM absorption can cause the centroid of Ly$\alpha$ to be artificially redshifted (i.e.
other lines blueshifted relative to Ly$\alpha$) up
to about 100--200~km/s; therefore,
blueshifted velocities of \cii\ line up to $\Delta v = v_{\rm [C\textsc{ii}]} - v_{\rm Ly\alpha} = -200$ km/s  are 
fully consistent with the rest-frame of the galaxy.
Positive velocity shifts and velocity shifts larger than 300~km/s can be  explained invoking outflows, often
traced by Ly$\alpha$ and high ionisation lines, and higher column density of neutral hydrogen in the galactic or
circumgalactic medium.
However, IGM absorption and outflows cannot account for velocity shifts relative to other UV transitions (e.g. CIII or
UV absorption features, as in the case of \citealt{Capak:2015}) and obviously cannot account for velocity shifts
between different far-IR transitions, as observed by us and others \citep{Pavesi:2016, Decarli:2014}. 
In fact, all far-IR and UV emission lines, whose emission is not absorbed by IGM, should have the same velocity offsets if they are emitted from the same region of the galaxy.

The lack of any clear correlation in Fig.~\ref{fig:offset} with the SFR and the spatial offsets larger then few kpc may yield to a further  interpretation.
Major merging, minor merging (i.e. accretion phases) and/or virial motions
of different galaxy components in the early phases of formation, subject to different excitation mechanisms, different
chemical abundances and also differential extinction effects, are likely to play a major role in the observability
of different spectroscopic tracers and on their distribution and offsets. This will be discussed further in the next
section.

Summarising, while some of the spatial offsets may be ascribed to positional uncertainty and some of the
spectral offsets may be ascribed to Ly$\alpha$ IGM absorption and/or outflows, a fraction of the offsets
observed between optical emission and far-IR emission (and also between different far-IR tracers)
may be due to physically distinct regions of a galaxy, characterised by different physical conditions,
or accretion events (minor merging) or major merging.

%Displaced \cii\  emission is indeed predicted by recent models of primeval galaxies in the re-ionisation epoch \citep{Salvadori:2009, Vallini:2013, Dayal:2014, Graziani:2015, Vallini:2015, Vallini:2016}. 
%According to \cite{Vallini:2015,Vallini:2016}, strong stellar feedbacks and photo-evaporation ionise and destroy molecular clouds, hence suppressing the emission of \cii\  from PDRs. 
%However, there can be clumps of neutral gas at galaxy outskirts, in the process of accreting onto the central object. These clumps can survive photoionisation and in turn shine in \cii\ and \oiii, as a consequence of the diffuse Far-UV radiation (FUV Habing band: 6 eV $<h\nu<$ 13.6 eV) emitted by the primary galaxy.
%Other models simply expect that displaced emission traces metal-enriched dust-obscured systems that are not
%detected at UV rest-frame wavelengths, while the UV emission traces metal-poor involved systems
%slightly obscured by dust \citep[Katz et al., in prep;][]{Willott:2015}. 
%A more detailed analysis of these possibility will be
%explored for the specific case of \bdf\ in the following.

\subsection{Theoretical scenarios and numerical simulations}

Before discussing in detail our ALMA results on BDF-3299,
in this section we briefly review the numerical simulations that have been developed
in the past few years specifically focussed on the distribution of far-IR lines emission
in primeval galaxies, so that these provide possible scenarios in which our data can be interpreted.

\cite{Vallini:2013} take advantage of cosmological SPH simulations, combined with a radiative transfer code,
to investigate the distribution of the various phases of the ISM and CGM in a primeval galaxy at z$\sim$6.6.
They predict that the bulk of the galaxy is mostly photoionised by the strong UV radiation
field of young stars, and they expected tracers of ionised gas (including [NII]) to be co-located with
the regions of star formation. In their simulation \cii\ is expected to survive only in (accreting) clumps
of pre-enriched gas in the circumgalactic medium; in these clumps the \cii\ emission results as a consequence
of excitation by the UV radiation received externally from the central galaxy (i.e. not as a consequence of
in-situ excitation by in-situ star formation). In this scenario, offset of \cii\
emission relative the UV counterpart (star forming galaxy) results naturally. However, \oiii\ should be colocated
with the UV counterpart.

\cite{Vallini:2015} and \cite{Pallottini:2016} expand further the model by
including \emph{(i)} the \cii\ emission from dense PDR surrounding molecular clouds, \emph{(ii)} a proper
modelling of the metallicity profile in the simulated galaxy \citep{Pallottini:2015} and, \emph{(iii)} by taking into account the excitation by the CMB.
%\cite{Vallini:2016} and \cite{Pallottini:2016} expand further the simulation by properly including feedback,
%following the metal enrichment, the formation of molecular clouds, and by also taking into account the excitation by
%the CMB. 
In these simulations, although circumgalactic \cii\ is present, the peak of the \cii\ emission
is coincident with the location of molecular clouds, where stars form. In this scenario, although a component on
large scales is present, the \cii\ emission peaks are expected
to trace regions of star formation, which should be coincident with the Y-band emission, unless dust obscuration plays
a role. \cite{Vallini:2016} suggests that an offset between \cii\ emission and UV-stellar emission could result
from a strong feedback effect, which could clean the most vigorous star forming regions, or even the entire galaxy,
from the bulk of the molecular gas. 
Low metallicity and chemical inhomogeneity are also invoked as possible causes of the faintness of the \cii\
emission; indeed, metallicity effects were already proposed in \cite{Vallini:2015} to be one of the major cause of
the deficit of the \cii-SFR relation observed at high-redshift relative to the local one \citep[e.g.][]{De-Looze:2014}.
%Low metallicity and chemical inhomogeneity also are invoked as possible
%causes of the faintness of the \cii\ emission.
\cite{Vallini:2016} also point out that the relative intensity of the \oiii\ and
\cii\ emission strongly depends on the intensity of the radiation field $G_0$, on the gas metallicity and
on the evolutionary stage of the system.

\cite{Katz:2016} present six cosmological hydrodynamic simulations, including radiative transfer, and investigate
the evolution of star formation, chemical enrichment and the strength and hardness of the radiation field,
as well as the distribution of the various gas phases (and in
particular the species responsible for the emission at far-IR wavelengths) on different galactic and circumgalactic
scales. They find that \cii\ has a diffuse component, but its most prominent peaks are associated with dense clumps
in which star formation is taking place. \oiii\ traces gas ionised in star forming regions, but predominantly
traces those regions in which the ionisation parameter is highest. They do observe offsets between
\cii\ and \oiii, primarily consequence of variations of the ionisation parameter. According to their simulations
regions of star formation are always associated with either \cii\ clumps or \oiii\ clumps. They suggest that offsets
of the \cii\ and/or \oiii\ emission relative to the UV emission are primarily associated with dust extinction
and/or metallicity effects. In particular, they suggest that the absence of UV counterparts in regions emitting
\cii\ and/or \oiii\ is a consequence of dust extinction. Instead, the weakness or absence of \cii\ or \oiii\ emission
in some star forming regions visibile at UV wavelengths is a consequence of low metallicity; more specifically,
these could be low metallicity systems recently accreted through the cosmic streams, whose low dust content makes them
visibile in the UV.

In the following we discuss and interpret the data specific of BDF-3299, also at the light of the scenarios
discussed above.

\subsection{The lack of far-IR emission at the UV location of BDF--3299}

Although some marginal \cii\ emission is detected at the location of the Y-band counterpart (tracing
the UV rest-frame light) of \bdf,
and although the 2$\sigma$ emissivity contour of \oiii\ does overlap with the Y-band image,
both far-IR tracers peak at a few kpc away from the UV emission.

Since both \cii\ and \oiii\ are only marginally detected at the location of the Y-band counterpart,
we assume that neither of them is detected and we infer their upper limit.
The upper limit on \cii\ at the location of the UV emission was already inferred by
\cite{Maiolino:2015} and estimated to be \lcii$<2\times10^{7}$ \lsun. 
To estimate the upper limit on \oiii\, and
to be consistent with the upper limit on \cii\ luminosity inferred by \cite{Maiolino:2015}, we
assume a line width of 100 km/s; by doing so we obtain an upper limit on the
\oiii\ luminosity of \loiii$<5.5\times10^{7}$ \lsun.

As mentioned above, the
lack of \oiii\ and \cii\ emission at the location of the UV-emission tracing young stars, could be associated with low metallicity in the primeval galaxy.
Using the {\sc cloudy} code, \cite{Inoue:2014} predict the  FIR-line emissivity in HII regions from high-$z$ galaxies as a function of metallicity ($Z$). 
They found that the \oiii\ emissivity is proportional to the metallicity at $Z<Z_\odot$, while the dependence
becomes shallow at  $Z>Z_\odot$.
The model predicts that  a galaxy with an \oiii\ luminosity of $5.5\times10^7$ \lsun\ and a SFR=7 \sfr\
(inferred by the UV emission) can be explained with a metallicity in the range
of $Z=0.1-0.3~Z_{\odot}$, depending on ionisation parameter ($-3.0<\log_{10}U<-1.0$) and on the hydrogen number density  ($0.0<\log_{10}(n_{\rm H}/{\rm cm}^{-3})<3.0$).
Hence, according to these models, a metallicity lower than $Z<0.1~Z_{\odot}$ should result into
an \oiii\ luminosity undetectable by our observations. Such low metallicity would be consistent with the low
dust content discussed in \cite{Maiolino:2015}.

An alternative scenario, already mentioned in \cite{Maiolino:2015} and in \cite{Vallini:2015, Vallini:2016},
is that we are observing \bdf\
at the end of a strong feedback phase that swept away the dense gas out of the galaxy,
hence suppressing the emission of far-IR lines. Ly$\alpha$ would
still be detectable, despite the low gas content; indeed, Ly$\alpha$ has little dependence on the total amount of gas since it is mostly sensitive to the UV ionising flux and to the escape fraction, which might be enanched during a strong feedback event. 
%,since this line is mostly an ionising photon counter,
%hence its intensity is little sensitive to the total amount of gas, being mostly sensitive to the
%UV ionising flux and to the escape fraction.
This scenario is discussed in \cite{Vallini:2013, Vallini:2015, Vallini:2016} and \cite{Olsen:2015}. 

 By using the model by \citet[][\citetalias{Vallini:2016} hereafter]{Vallini:2016}, we can quantify the maximum gas mass contained in \bdf\ 
 that  can be consistent with the lack of detection.
\citetalias{Vallini:2016} model the FIR emission from molecular clouds (MC) immersed in the radiation field of a parent galaxy. The model
assumes that no star formation activity is present inside the clouds and accounts for photoevaporation effect due to
external radiation.  In \citetalias{Vallini:2016} the MC internal structure is accounted for by considering the
co-existence of different density phases ($10< n_{\rm H}/{\rm cm}^{-3}< 10^5$), that are due to competition between
turbulence support and gravitational collapse \citep[e.g.][]{Federrath:2013}. For the present work, we treat the FIR
emitting gas as collection of MCs with total mass \mgas\ and, as a baseline, we assume a gas metallicity
%\footnote{Assuming a different gas metallicity does not hinder the conclusions of our analysis.}
$Z=0.2\,Z_{\odot}$ (consistent with the predictions by \cite{Inoue:2014}).
In Fig.\ref{fig:model} we plot the predicted FIR emissions of \cii\ and \oiii\ as a function of incident flux (G) for different \mgas.
%\ (with $Z=0.2\,Z_{\odot}$).
% Thus we can calculate the FIR emission as a function of the incident flux and the mass of the gas.
%Note that we have assumed that star formation has been ongoing for 100 Myr.
The  FUV density field in \bdf\ can be estimated by using {STARBURST99} \citep{Leitherer:1999}. Based on
the UV photometric constraints \citep{Vanzella:2011}, the galaxy has a luminosity in the FUV
band of $\sim 3\times10^{43}{\rm erg}\,{\rm s}^{-1}$. 
To derive the size of the star-forming galaxy, we fit a two dimensional Gaussian to the Y-band and we estimate a
size of $\sim1\, {\rm kpc}$ for \bdf , which implies a mean
radiation field in the galaxy of $\sim 10^3\, G_{0}$ ($G_{0}=1.6\times10^{-3} {\rm erg}\, {\rm cm}^{-2}\, {\rm s}^{-1}$,
\cite{Habing:1968}).
This value is plotted as the rightmost vertical line in Fig.~\ref{fig:model}.
We note that such a value must be taken only as an average reference, as inside a galaxy the FUV flux should be
calculated accounting for the distribution of stars and gas \citep{Wolfire:2003}.

 The upper limits on \lcii\ an \loiii \ 
inferred for \bdf\ are marked with horizontal green dashed lines. 
Since the \cii\ and the \oiii\ emission are not detected at the location of the UV-emission, we conclude that the gas
mass in \bdf\ is less than $10^{9}$ \msun. Of course, a different metallicity  would imply a different upper limit
on the gas mass (approximately scaling with the metallicity, for example see eq.~8 in \citealt{Pallottini:2016}).

We also note that BDF-3299 is not detected in the continuum emission at $\rm \lambda _{rest}=88 \mu m$.
If a  standard SED of a starburst galaxy (M82) is adapted to the observed upper limit on the continuum
($\rm F_\nu < 40\mu Jy$) then the inferred (2$\sigma$) upper limit on the SFR would be 12\sfr, consistent with
the SFR inferred from the UV light. However, if the dust content is much lower than in local starburst templates,
as a consequence of low metallicity
and because of feedback removal of the ISM, then even less radiation is expected to be reprocessed in the far-IR
and the constraint on the SFR would be even looser.

%
%%%%%%%%%%%%%%%%
%The expected flux from \bdf\ on the \clumpa\ and clump-I is then calculated by using the observed projected distance.
%For clarity sake, we assume equal distances for the \oiii\ and \cii clumps, i.e. $\simeq 3.5 {\rm kpc}$. Hence, we obtain a flux of $\sim 150\, G_{0}$. Note that if projection or absorption effects are in place, the flux from the galaxy would be reduced.

%According to these scenarios, both \oiii\ and \cii\ emission may arise from surrounding satellite clumps
%heated by  the strong UV radiation from the star-forming galaxy. This will be discussed further in the
%next sections.  
%%%%%%%%%%%%%%%%

\subsection{The nature of the \cii\ and \oiii\ emitting regions}

Two main scenarios can be envisaged for explaining the offset \cii\ and \oiii\ emission relative to the
UV emitting region. One is that, as
mentioned above, these are accreting satellite clumps or clumps resulting from past outflow, and which are
being excited by the UV radiation emitted by \bdf. The second scenario is that the line emission in 
these clumps is due excitation by  in-situ star formation that is dust-obscured, hence not detected
at UV-rest frame wavelengths. In the latter case these star forming regions could be part of the same large galaxy,
but subject to differential dust extinction and differential excitation and enrichment effects, or could be
galaxies in the process of (major) merging and/or accreting small galaxies (minor merging).

These scenarios can be investigated by using the same \citetalias{Vallini:2016} models discussed above. To investigate the
scenario of clouds being excited by the UV radiation emitted by BDF-3299 we estimate
the UV flux impinging
onto the  \oiii\ and \cii\ clumps by using the observed projected distance.
For sake of simplicity and clarity, we assume an equal distance
of $\simeq 3.5 {\rm kpc}$ for both the \oiii\ and \cii\ clumps with respect to BDF-3299.
Hence, we obtain an intensity of the UV radiation field of $\sim 150\, G_{0}$.
We note that the latter is an upper limit, since
if projection or absorption effects are in place then the resulting UV flux from BDF-3299 would be reduced
consequently. The inferred Habing field is indicated by the leftmost vertical line in
Fig.~\ref{fig:model}. The \cii\ and \oiii\ observed luminosities are indicated with
horizontal lines.

\subsubsection{Oxigen emission}
Clearly the observed \oiii\ luminosity cannot be explained in terms of external excitation from the UV radiation
emitted by \bdf\, as the implied gas mass would have to be much larger than $10^{10}~M_{\odot}$. Therefore,
the only viable explanation is that the \oiii\ is powered by in-situ star formation.
This is consistent with the analysis of \citetalias{Vallini:2016}, where it is shown that the typical current ALMA sensitivity is not sufficient to detect \oiii\ emission in clumps located a few kpc from the main galaxy with a radiation field similar to that produced by only \bdf.
%This is consistent with the finding of \citetalias{Vallini:2016}, who find that the current ALMA sensitivity is not sufficient to detect
%\oiii\ emission in clumps located a few kpc from the main galaxy and immersed in the radiation field similar to that
%emitted by \bdf.  
Therefore, the detection in the clump-I indicates that \oiii\ excitation primarily originates from
in-situ star formation.

If we take the relation between SFR and \oiii\ luminosity for metal poor galaxies found by \cite{De-Looze:2014},
then the \oiii\ luminosity observed in Clump-I  yields a SFR of about 7.5~\sfr. 

If unobscured, this should result in about the same Y-band flux as BDF3299. However, at the location of Clump-I
there is no Y-band detection, with an upper limit on the flux that is about seven times lower than the flux
observed in BDF-3299.
The tension can be reconciled by assuming a mild amount of dust extinction. As a matter of fact,  at the rest-frame wavelength corresponding to the Y-band image (i.e. $\sim$1300\AA)  and assuming a Calzetti attenuation curve \citep{Calzetti:2000}, a reddening of only E(B-V)=0.2 (i.e. $\rm A_V\sim 0.8$) would be enough to make the UV emission undetectable.
In fact, with such extinction we predict a magnitude in Y-band $>30$, while the magnitude limit of the current HST observations is 28 at 10$\sigma$).
We note that a similar reddening value ($<E(B-V)>\sim0.19$) has been recently observed in $z\sim1$ low-metallicity ($Z<0.6Z_\odot$) star-forming galaxies \citep{Amorin:2015}. 

We shall also mention that the SFR of about 7.5~\sfr inferred from the \oiii\ luminosity is compliant with
the non-detection of continuum dust emission. 
In fact, the flux emission at 230 GHz expected by a high-$z$ star-forming galaxy with SFR=7.5 \sfr\ is $\sim5$ \ujy\ (even fainter in the low-metallicity scenario) that is below  our current sensitivity. 
 Additionally, if the in-situ star forming gas in the clump is very concentrated, then
the G flux would be even higher, and lower values of gas mass (and consequently SFR) are needed in order to explain the \oiii\ emission, as shown in Fig. 6.

We also note that the non-detection of \cii\ at the location of clump-I implies $\rm L_{[O{\sc III}]}/L_{[C{\sc II}]}>8$, which is
is consistent with the value expected for low metallicity galaxies and with the \oiii /\cii\ ratio observed by Inoue et al.
(2016) in a galaxy at z$=$7.2.

The high value of the \oiii /\cii\ ratio can also be explained in terms of high ionisation parameter, as expected
by \cite{Katz:2016}.

It should be noted that also the other two \oiii\ detections (clumps II and III), which are even farther away
from BDF-3299, can only be explained in terms of in-situ star formation (with star formation rates
of about 10--20 \sfr). Also for these cases the lack of Y-band counterpart can be accounted for in terms of some dust
reddening corresponding to E(B-V) less than 0.3,  which is consistent, within the errors, to the dust reddening estimates reported by \cite{Amorin:2015} who observed  $z\sim1$ low-metallicity ($Z<0.6Z_\odot$) star-forming galaxies with 0.1\sfr$<$SFR$<$35\sfr.

%It is worth noticing that targeting \oiii\88$\mu$m is a powerful tool to detect star forming galaxies
%at z$>$7 with modest SFR, competitive with the HST sensitivity, with the additional advantage of
%being insensitive to dust extinction.

\subsubsection{Carbon emission}
\begin{figure*}
\centering
 \includegraphics[width=1.5\columnwidth]{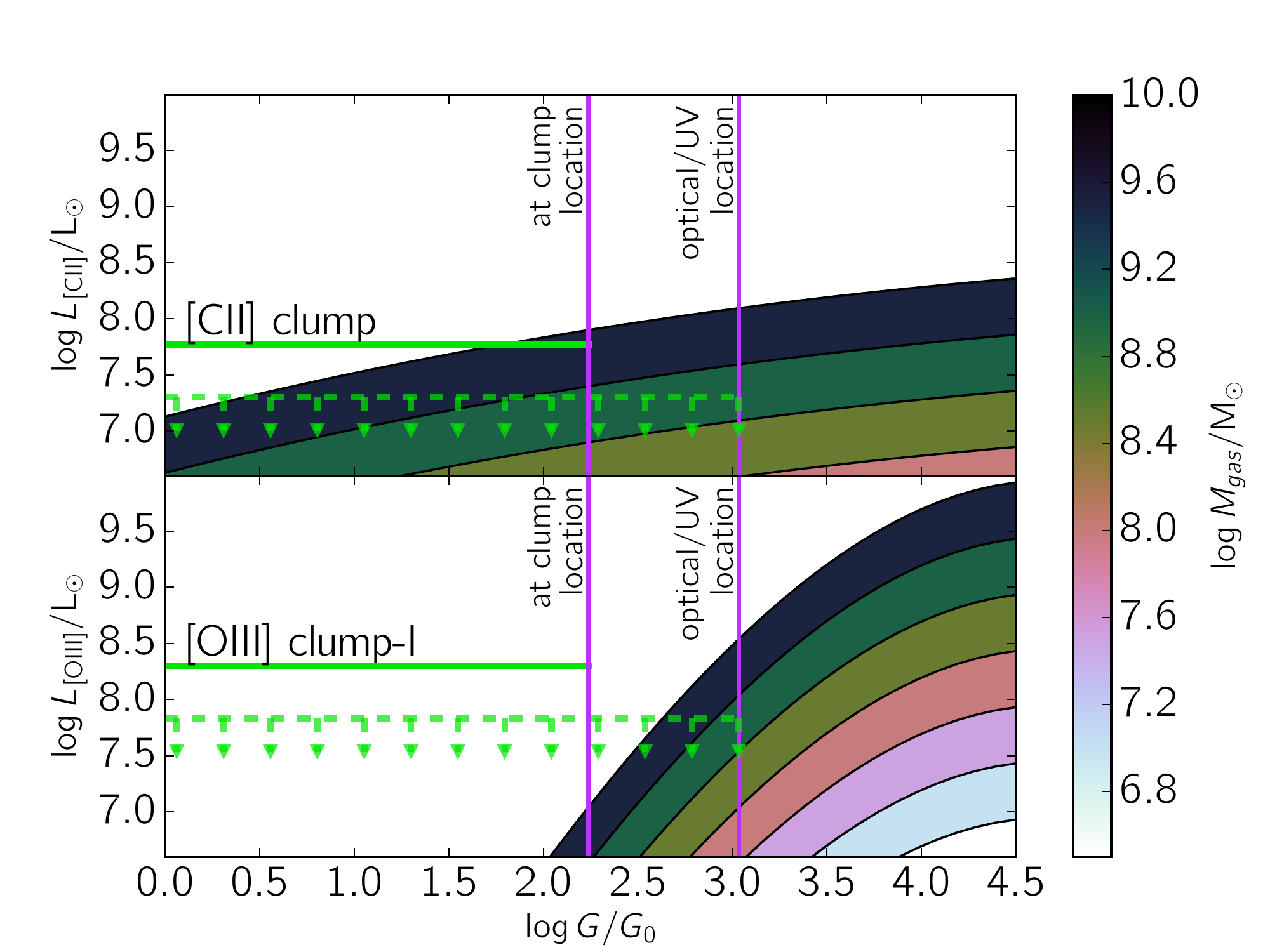}%,trim={0 7cm 0 6cm},clip]
 \caption{Predicted \cii\ (upper panel) and \oiii\ (lower panel) luminosities ($L$) as a function of the incident flux ($G$) exciting the line, inferred
 by using the model presented in \citetalias{Vallini:2016}, in which the colour shading is done according to $M_{gas}$, as encoded in the colourbar.
 In each panel we plot with a horizontal green line
 the luminosity inferred from our observations of the 
\cii\ clump and of the \oiii\ clump-I. The horizontal dashed green lines show the upper limits at the location of the Y-band image of BDF-3299. 
Additionally, we show the flux $G$ expected from \bdf\ at the location of the \cii\ and \oiii\ clumps and also the estimated average $G$ within
the galaxy BDF3299. See the text for details.
 }
 \label{fig:model}
\end{figure*}

Regarding the \cii\ emission clump, Fig.~\ref{fig:model} illustrates that a satellite system with total mass of
about $\rm 4\times 10^9~M_{\odot}$ would radiate the observed \cii\ if excited externally by the UV light
emitted by BDF-3299. Of course, the limiting
mass depends significantly on the assumed metallicity: a metallicity higher than assumed in  Fig.\ref{fig:model}
(i.e. $\rm Z>0.2 ~Z_{\odot}$), would relax the gas mass requirements; higher metallicity could
result from pre-enrichment by past outflows
from BDF-3299 or from some enrichment due to past minor star formation in the cloud, as discussed
in \cite{Maiolino:2015} . In summary, the possibility of a satellite clump (either in the process of accreting,
or resulting from previous outflow) whose \cii\ emission is excited by the UV radiation emitted by \bdf\
is viable.

The alternative scenario of excitation by in-situ star formation cannot be excluded either.
Indeed, according to
the \cii-SFR correlation for low metallicity galaxies inferred by \cite{De-Looze:2014}, the observed
\cii\ luminosity would imply a SFR of only 3 \sfr. The latter would be only marginally detected in the Y-band
($<3\sigma$) and a reddening of only E(B-V)$=$0.1 (i.e. $\rm A_V\sim 0.4$) would be certainly enough to make
it completely undetectable.

In the latter scenario the absence of \oiii\ emission can be interpreted in terms of metallicity  (higher
metallicity associated with the \cii\ clump than in the \oiii\ clump), or in terms of lower ionisation parameter
than in the \oiii\ clump, as suggested by \cite{Katz:2016}.
The latter effect has already been observed in some SMGs
at high-z \citep[e.g.][]{Pavesi:2016,Rawle:2014a,Riechers:2014} in which different far-IR lines show significant
spatial and velocity
offsets. Our result may imply that
a similar scenario applies also to normal galaxies at high redshift.

To further discriminate these different scenarios higher sensitivity \cii\ and \oiii\
ALMA data are required, possibly also including
[NII]122$\mu$m, which would enable to better disentangle excitation and metallicity gradients. JWST will
also greatly help by targeting the nebular lines and stellar continuum with much higher sensitivity than
currently feasible and at optical rest-frame wavelengths, at which dust absorption is much reduced relative
to the optical.

\section{Summary}
\label{sec:summary}

We have reported new ALMA millimetre observations aimed at investigating the \cii\ at 158$\mu$m and the \oiii\ at
88$\mu$m emission in the star-forming (SFR=5.7 \sfr) galaxy  \bdf \ at z=7.1.  
The goal of this work has been to 1) obtain a detailed map of the cold, clumpy gas structure by analysing a new \cii\
observations at a resolution significantly higher (a factor $\sim 3$) than previous observations \citep{Maiolino:2015};
2) map the \oiii\ emission in this primeval galaxy in order to further constrain the nature of its nature. 

About 70\% ($\pm$20\%) of the \cii\ emission detected by \cite{Maiolino:2015} is resolved out in the high-angular resolution
images indicating that a significant fraction of the \cii\ emission is extended and diffuse on scales larger than
about 1~kpc. We have combined the old ALMA low resolution  and the new ALMA high resolution data
to obtain deeper images at intermediate angular resolution. The resulting data show that the \cii\ emission
has a complex, clumpy structure. Both diffuse emission and multi-clumpy structure of the of the cold gas
traced by the \cii\ emission, in and around
primeval galaxies, is expected by recent models of primeval galaxies
\cite[e.g][]{Vallini:2013, Vallini:2015, Vallini:2016,Pallottini:2015, Pallottini:2016,Fiacconi:2016,Katz:2016}.

The new combined datacube also confirms the offset of \cii\ relative to the optical (Y-band) counterpart, by
about 0.7$''$ (4~kpc), although marginal detection of \cii\ is also found at the location of the optical counterpart.

%In this case, we detect the \cii\ emission with a level of confidence of 6.2$\sigma$. 
%However, the flux map obtained from the combined datasets is slightly different from that show in \cite{Maiolino:2015}. 
%In fact the \cii\ map exhibits structures smaller than beam size indicating a multi-clump morphology. 
%In addition the \cii\ line profile  has a line width narrower than that  measured by \cite{Maiolino:2015} indicating that the observed emission line is the contribute of multiple narrow \cii\ emission lines due to the presence of several emitting cold neutral molecular clumps, as expected by models \cite[e.g][]{Vallini:2013}.

Furthermore we have reported the detection of 
\oiii\ emission line, with a level of confidence of 7.3$\sigma$. The line is also offset relative to the
optical counterpart, in the same direction as the \cii\ emission. However, most of the \oiii\ emission does
not overlap with the \cii\ emission.
Two additional \oiii\ line emitters are also identified in the ALMA field of view, within 20 kpc of \bdf.

We have shortly reviewed the 
the velocity and position of far-IR line emission in high-$z$ systems reported in literature, and we have shown that
offsets relative to the optical counterparts, spatially or in velocity, are relatively common.
We find that the offset, either spatial or in velocity, does not correlate with the SFR.
While in
some cases these offsets
may be ascribed to astrometric uncertainties and to IGM absorption of Ly$\alpha$, in other
cases the offset are certainly associated with physically distinct components traced by optical and
far-IR emission.

By comparing our data with models we show that the observed
\oiii\ emission cannot be excited by the UV radiation coming
from BDF-3299. The \oiii\ emission must be excited by in-situ star formation, with a SFR$\sim$7\sfr.
The lack of Y-band (UV-rest frame) counterpart at the location of the \oiii\ clump(s) can be explained
in terms of some modest amount of dust extinction.
We note that \oiii88$\mu$m can be a powerful tool to
detect normal star-forming galaxies with ALMA at z$>$7, competitive with HST, with the additional
advantage of being insensitive to dust extinction.

We note that the non-detection of \cii\ at the location of the \oiii\ clumps is consistent with the lower
limit on \oiii/\cii\ observed with ALMA in a similar high-z galaxy at z=7.2, and with local low metallicity galaxies.
However, the high values of  \oiii/\cii\ can also be interpreted in terms of high ionisation parameter,
as suggested by some models.

The origin of the \cii\ emission is more uncertain. Diffuse gas, associated with either satellite (accreting)
gas rich systems or gas expelled by the optical galaxies, excited by the UV radiation field emitted
by BDF-3299, is a viable interpretation based on the comparison with models. However, the scenario
in which the bulk of the \cii\ clumpy emission is excited by in-situ star formation (at a level of 3~\sfr) cannot
be excluded either. The faint/absent \oiii\ emission at the location of \cii\ can be interpreted in terms
of either moderately high metallicity or low ionisation parameter.

Overall, the observational properties can be interpreted 
as a primeval system in the process of being assembled,
in which different emission lines trace different components, characterised by different metallicity
or excitation (ionisation parameter) properties. In this system the observable UV light is associated with
the least obscured region, either because recently accreted, hence with low metallicity, or because strong
feedback has removed the bulk of dust and gas content; both scenarios would account for the weakness of far-IR
line emission at the location of the optical image.

Similar properties are
observed in the much more powerful SMG systems at high redshift. Our observations are revealing that
similar complex structures and processes are occurring also in normal primeval galaxies, with modest
SFR.

Deeper ALMA data, possibly targeting also other transitions,
are needed to obtain more detailed description of the system and to discriminate between different viable scenarios.
JWST will also greatly help in understanding the nature of these system by probing the stellar population at
optical rest-frame wavelengths and various nebular optical lines.

\begin{acknowledgements}
This paper makes use of the following ALMA data: ADS/JAO.ALMA\#2012.1.00719.S, ADS/JAO.ALMA\#2012.A.00040.S and,ADS/JAO.ALMA\#2013.A.00433.S ; which can be retrieved from the ALMA data archive: https://almascience.eso.org/ alma-data/archive. ALMA is a partnership of ESO (representing its member states), NSF (USA) and NINS (Japan), together with NRC (Canada) and NSC and ASIAA (Taiwan), in cooperation with the Republic of Chile. The Joint ALMA Observatory is operated by ESO, AUI/NRAO and NAOJ.
Based on observations made with the NASA/ ESA Hubble Space Telescope , obtained at the Space Telescope Science Institute, which is operated by the Association of Universities for Research in Astronomy, Inc., under NASA contract NAS 5-26555. These observations are associated with program \# 13688.
We thank Harley Katz and Martin Haehnelt for comments and discussions.
SC acknowledges support by the Science and Technology Facilities Council (STFC).
RM acknowledges support by the Science and Technology Facilities Council (STFC) and the ERC Advanced Grant 695671. 
``QUENCH''.
\end{acknowledgements}
%%%%%%%%%%%%%%%%%%%%%%%%%%%%%%%%%%%%%%%%%%%%%%%%%%



\bibliographystyle{aa} % style aa.bst
\bibliography{bibliography_bdf3299} % if your bibtex file is called example.bib


%%%%%%%%%%%%%%%%%%%%%%%%%%%%%%%%%%%%%%%%%%%%%%%%%%

%%%%%%%%%%%%%%%%% APPENDICES %%%%%%%%%%%%%%%%%%%%%

\appendix

\section{Extended \cii\ emission}\label{sec:appa}
 
In this appendix, we compare the \cii\ emission detected in the ALMA Cycle 1 and Cycle 2 and we discuss the differences between the outcomes of these projects.

\begin{figure}
\quad \quad
\includegraphics[width=0.7\columnwidth]{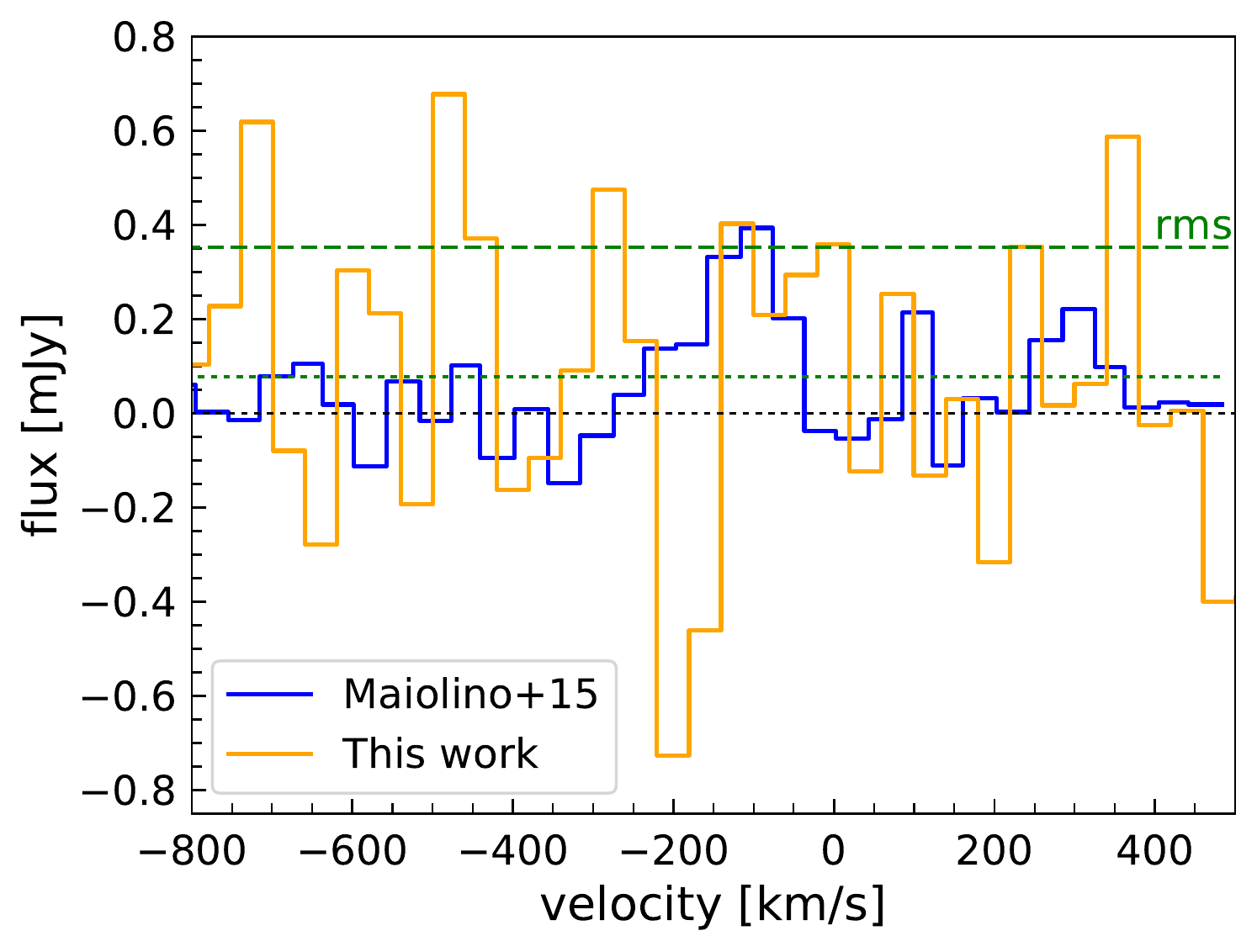}
  \includegraphics[width=0.7\columnwidth]{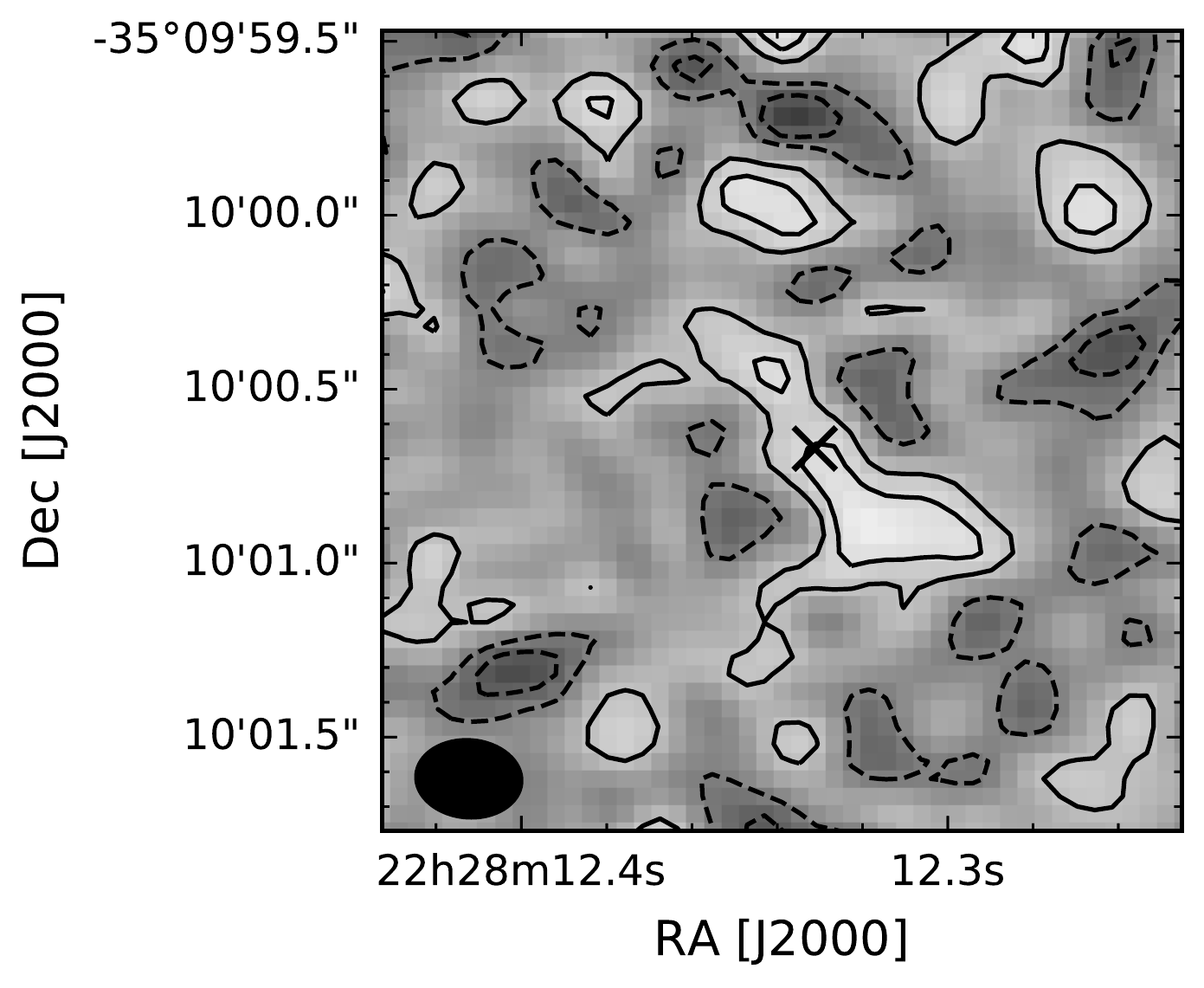}
 \caption{
 Top: ALMA spectrum of the  \cii\ detections, extracted from the an elliptical aperture of size 0.55\arcsec$\times$1.03\arcsec\ with resolution of 40 km/s. The blue solid line is the spectrum  by \cite{Maiolino:2015}, while the orange and green lines show the  spectrum and  noise level    obtained from the new high-resolution observations, respectively. 
Bottom: High-resolution ALMA map (angular-resolution = 0.3\arcsec$\times$0.2\arcsec) obtained by integrating the line under the gold shaded region in the top panel of Figure~\ref{fig:cii}a (i.e. -70$<v<10 $ km/s).
%with a velocity interval of 100 km/s, centered at –64 km/s relative to Ly$\alpha$ redshift. 
Black contours are at levels of -2,-1,1, and 2 times the noise per beam in the same map, i.e. 9 mJy/beam km/s. }
 \label{fig:cycle2}
\end{figure}

The Top panel of Figure \ref{fig:cycle2} shows the Cycle 2 ALMA spectrum  extracted at the location of the  \cii\ detection (RA(J2000)=22:28:12.325, DEC(J2000)=-35:10:00.64) and from a elliptical aperture of 0.55\arcsec$\times$1.03\arcsec  (approx. nine beams) corresponding to the extension of the clump inferred by \cite{Maiolino:2015}. The carbon line is not detectable in the integrated line because of  the low sensitivity in the high-spatial resolution observations. In fact, the noise level (0.35 mJy in element channels of 40 km/s) in the integrated spectra is comparable with the peak of \cii\ line (0.39 mJy). 
The high-resolution integrated map (bottom panel of Figure \ref{fig:cycle2}), whose sensitivity is 1.4 times lower than the integrated map obtained with the compact ALMA configuration,  does not reveal any high-significance extended or multi-clumpy emission, suggesting that the \cii\ emission is mostly resolved out in the ALMA extended configuration, although there a number of 1$\sigma$ and 2$\sigma$ emissions spots that will contribute to the signal in the total, combined data set. In fact, if the emission was powered by a compact source unresolved at this resolution, we would expect to observe the peak of the emission with a level of confidence of $\sim3-3.5$ in the integrated map.

%%%%%%%%%%%%%%%%%%%%%%%%%%%%%%%%%%%%%%%%%%%%%%%%%%

\end{document}